\documentclass[twocolumn,floatfix,aps]{revtex4-1}
\usepackage{bm}% bold math
\usepackage{graphicx,color}
\usepackage{graphicx}% Include figure files
\usepackage{dcolumn}% Align table columns on decimal point
\usepackage{times}

\usepackage{amsmath,amssymb}

\begin{document}

\title{Exact solutions of the harmonically confined Vicsek model} 
\author{L. L. Bonilla$^*$}
\affiliation{Universidad Carlos III de Madrid, Gregorio Mill\'an Institute for Fluid Dynamics, Nanoscience and Industrial Mathematics, Avenida de la Universidad 30; 28911 Legan\'{e}s, Spain}
\affiliation{Universidad Carlos III de Madrid, Department of Mathematics, Avenida de la Universidad 30;  28911 Legan\'{e}s, Spain. 
$^*$Corresponding author. E-mail: bonilla@ing.uc3m.es}
\author{R. Gonz\'alez-Albaladejo}
\affiliation{Gregorio Mill\'an Institute for Fluid Dynamics, Nanoscience and Industrial Mathematics, Universidad Carlos III de Madrid, 28911 Legan\'{e}s, Spain}

\date{\today}
\begin{abstract}
The discrete time Vicsek model confined by a harmonic potential explains many aspects of swarm formation in insects. We have found exact solutions of this model without alignment noise in two or three dimensions. They are periodic or quasiperiodic (invariant circle) solutions with positions on a circular orbit or on several concentric orbits and exist for quantized values of the confinement. There are period 2 and period 4 solutions on a line for a range of confinement strengths and period 4 solutions on a rhombus. These solutions may have polarization one, although there are partially ordered period 4 solutions and totally disordered (zero polarization) period 2 solutions. We have explored the linear stability of the exact solutions in two dimensions using the Floquet theorem and verified the stability assignements by direct numerical simulations.
\end{abstract}
\maketitle
%\begin{quotation}
%\end{quotation}

\section{Introduction}\label{sec:1}

The formation and evolution of flocks are important topics in biological physics, soft and active matter and complex phenomena in social sciences. In the past thirty years, advances in quantitative observation and theory have given renewed impulse to this field. In particular, ideas from phase transitions, stochastic thermodynamics and statistical mechanics applied to idealized models have had an enormous impact. 

Consider for example the Vicsek model (VM) \cite{vic95}, paradigmatic in the study of {\em dry active matter} \cite{cha20}. Individuals in a flock are point particles moving at constant speed that update their positions after a time $\Delta t$ by adopting as their new velocity vector that of the mean velocity of their neighbors plus an alignment noise (majority rule). Neighbors are defined by their metric or topological distance: all particles inside a sphere centered at a given particle (metric) or the $n$-closest particles (topological). The particles are confined in a box with periodic boundary conditions. That individuals are point particles indicates that we are interested in length scales larger than individual size. This is a simplification with a long tradition in physics: recall celestial bodies considered as point particles and the collapse of stars in statistical mechanics models \cite{gun07,alb20,chavanis20}. The VM discrete time dynamics is less usual but it indicates that the individuals need a certain time $\Delta t$ to measure their neighborhood and change their velocity accordingly. To avoid dispersion of the flock, we can either decree a confining periodic box or postulate some attraction force. The majority rule could play the role of confining force if all particles are supposed to be in the same neighborhood, which would produce a model of mean field type \cite{bir07}. In this case, the motion of the particles is either bounded and stationary or unbounded and migratory (all particles move together with constant velocity as in a fish school) \cite{bir07}.

The periodic VM has an ordering transition from a disordered gas phase of uniform density to ordered phases of moving bands and then to liquidlike polarized phases as the alignment noise decreases \cite{cha20}.  The moving bands on a periodic box are analogous to the migratory solutions for the VM that considers all particles to be in the same single neighborhood \cite{bir07}. The ordering transition is reminiscent of equilibrium phase transitions \cite{hua87,ami05} and therefore critical dynamics models with continuous time \cite{hoh77} have been proposed to described flocking formation, assumed to be an ordering transition. Among them the Toner-Tu model \cite{ton95}, its incompressible version \cite{che15,che18}, and the active version of Hohenberg-Halperin model G \cite{cav23}. There exist a well developed theoretical framework to study these models, including renormalization group and universality ideas \cite{ami05,wil83}. It is possible to calculate critical exponents using the renormalization group and claim that the dynamic critical exponent define a flocking dynamical universality class that may explain measured dynamic critical exponents  in swarms of midge flies \cite{cav23}, notwithstanding qualitative disagreements such as the shape of the swarm \cite{sin17} and quantitative ones such as the measured static critical exponents \cite{cav23,att14}. %advantage of these models 

To describe a particular flock we need to improve some of the less realistic features of the VM. Paramount among these are the periodic boundary conditions, which avoid particles to escape to infinity. We need to consider the biology of the flock \cite{oku86,par99,sum10,ton24} to impose a better boundary condition. Midge swarms form about salient features of the ground call markers and, unlike fish schools or bird flocks, do not migrate. For swarms, adding a confining harmonic potential to the VM makes sense because the corresponding linear spring force is compatible with the observed statistics of accelerations of individual midges in a swarm \cite{kel13}. Furthermore, the harmonic potential avoids particles escaping to infinity, which occurs if the periodic box of the standard Vicsek model is suppressed without including an attractive force \cite{gor16}. The resulting harmonically confined VM (HCVM) has interesting features including time periodic, quasiperiodic and chaotic phases, and a phase transition between chaotic and non-chaotic phases that occurs in three dimensional (3D) \cite{gon23,gon23mf,gon24} and two dimensional (2D) space \cite{gon232d}. For finitely many particles, there are different critical lines that form an extended criticality region. As the number of particles tends to infinity, these critical lines coalesce to zero confining potential at the same rate \cite{gon23,gon24}. These critical lines also coalesce as the alignment noise vanishes for finite particle number. Thus, the limit of vanishing noise can be used to unfold the phase transition \cite{gon24}. The static and dynamic critical exponents of the resulting scale-free-chaos phase transition agree with critical exponents measured in swarms of midges \cite{att14,cav23}. The HCVM also explains other qualitative features such as the swarm shape or the partial collapse of data for the dynamic correlation function at small scaled times \cite{gon23,gon24}.

%{\color{red} Explain mean field and exact solutions}
The HCVM is important for the application to insect swarms because its scale-free-chaos phase transition explains measured critical exponents and observed qualitative features. However, the evidence for the scale-free-chaos phase transition in the HCVM is numerical, mostly based in finite size scaling \cite{gon23,gon24,gon232d} and in mean field theory \cite{gon23mf}. Depending on the confinement and noise parameters, it is known that many particles may occupy the same sites \cite{gon23}. This is the basis of the mean field theory: The mean field HCVM assumes that all particles occupy the same single site at any given time \cite{gon23mf}. The continuous-time deterministic mean field VM with particles having an infinite region of influence has exact migratory and stationary solutions \cite{bir07}. The related Cucker-Smale model also has an infinite region of influence but it includes a weight function that decreases algebraically with the distance between particles \cite{cuc07}. It also has migratory configurations (asymptotic flocking) in which particles move with the same velocity  and keep fixed distances between them \cite{cuc07}. For this model and its variations including noise or time delay, different initial conditions produce solutions that evolve to asymptotic flocking \cite{cuc07,has13,erb16}. 

Do exact solutions exist for the HCVM? Allowing for groups of particles occupying the same sites, we have found families of time periodic and quasiperiodic {\em exact solutions} of the {\em deterministic} $d$ dimensional (dD) HCVM. While we have set $d=2$ to obtain the basic exact solutions, these solutions also solve the 3D (or dD) HCVM for planar configurations. We have found that the positions of many exact periodic and quasiperiodic solutions are on invariant circles. There are also period-2 and period-4 exact solutions whose positions are on a line or a rhombus. Within deterministic dynamics (HCVM without alignment noise) in dimension two,  we have found regions of the confinement parameter on which many exact solutions are stable, for the Floquet multipliers associated to equations linearized about them are not outside the unit circle. Our numerical simulations of the HCVM probe the nonlinear stability of the exact solutions and confirm the results of the linear stability analysis. 

The influence of alignment noise on the exact solutions is outside the scope of the present paper. From numerical simulations, we know that the critical exponents calculated from correlation functions are close to those found from the deterministic 3D HCVM and that, in the case of the noisy 3D HCVM, they can be extracted in the limit of vanishing noise \cite{gon24}. For the 2D HCVM, the location of the scale-free-chaos phase transition that separates chaotic and nonchaotic solutions at vanishing confinement has not been found: Numerical simulations always produce a positive largest Lyapunov exponent (indicating chaos; see \cite{gon23,gon24} for the algorithms we have used) that decrease as confinement values decrease down to the value beyond which the Benettin algorithm \cite{ben80} does not converge \cite{gon232d}. For the 3D HCVM with alignment noise, relevant chaotic attractors on the critical line appear through the quasiperiodic scenario \cite{gon23,gon23mf}. Thus, our exact 2D (or planar 3D) periodic and quasiperiodic solutions and further study of the small noise limit may contribute to the understanding of the scale-free-chaos phase transition.

This paper is organized as follows. The confined Vicsek model is presented in Section \ref{sec:2}. The basic periodic and quasiperiodic exact solutions are deduced in Section \ref{sec:3}. They include invariant circles (quasiperiodic solutions) and periodic solutions on a line and a rhombus. Periodic solutions with positions on a circle exist only for particular discrete values of the confinement whereas periodic solutions on a line exist for continuous values of the confinement larger than a certain lower bound. While these basic solutions have order parameter 1, Section \ref{sec:4} contains exact solutions that are disordered or partially ordered. Section \ref{sec:5} presents exact solutions corresponding to concentric orbits (either periodic or quasiperiodic solutions) that exist for appropriate confinements and remind of Bohr's old quantum theory. Linear stability of periodic solutions is considered in Section \ref{sec:6}, in which we also prove an extension of the Floquet theorem. Numerical results that shed light on the nonlinear stability of exact solutions are presented in Section \ref{sec:7}. Section \ref{sec:8} contains the conclusions of this work and the Appendix is devoted to technical considerations.

\section{Harmonically confined Vicsek model} \label{sec:2}
For finite $N$, the $d$ dimensional HCVM with parameters $\eta$, $\beta$ is:
\begin{subequations}\label{eq1}
\begin{eqnarray}
&&\mathbf{x}_i(t+1)=\mathbf{x}_i(t)+ \mathbf{v}_i(t+1),\quad i=1,\ldots,N,\label{eq1a}\\
&& \mathbf{v}_i(t+1)=v_0  \mathcal{R}_\eta\!\left[\Theta\!\left(\sum_{|\mathbf{x}_j-\mathbf{x}_i|<R_0}\mathbf{v}_j(t)-\beta\mathbf{x}_i(t)\right)\!\right]\!.\quad \label{eq1b}
\end{eqnarray}\end{subequations}
Here $\Theta(\mathbf{x})=\mathbf{x}/|\mathbf{x}|$, $R_0$ is the radius of the sphere of influence about particles, $v_0$ is the constant particle speed, $\beta$ is the confining spring constant, and $\mathcal{R}_\eta(\mathbf{w})$ performs a random rotation uniformly distributed on a spherical sector around $\mathbf{w}$ with maximum opening $\eta$ \cite{gon23}. Particles align their velocities with the mean of their neighbors within a sphere of influence except for an alignment noise of strength $\eta$. Eq.~\eqref{eq1} is the nondimensional version of the HCVM described in Appendix A of \cite{gon23}, where the resulting values of $v_0$ are of order one. Details about the algorithms used in numerical simulations of the 3D HCVM and calculations of the largest Lyapunov exponent to characterize parameter regions where the solutions are chaotic can be found in \cite{gon23}.

In 2D, Eqs~\eqref{eq1} can be written as
\begin{subequations}\label{eq2}
\begin{eqnarray}
&&\mathbf{x}_j(t\!+1)=\mathbf{x}_j(t)+ \mathbf{v}_j(t+1)\!,\quad  \label{eq2a}\\
&&\theta_j(t\!+1)\!=\!\mbox{Arg}\!\left(\sum_{|\mathbf{x}_k\!-\mathbf{x}_j|<R_0}\!\!e^{i\theta_j\!(t)}\!-\frac{\beta}{v_0} z_j(t)\!\right)\!\!+\!\xi_j(t),  \label{eq2b}\\
&&z_j=x_j+i y_j, \quad\mathbf{v}_j=v_0(\cos\theta_j,\sin\theta_j),\quad v_j=v_0e^{i\theta_j}\!.\quad\label{eq2c}
\end{eqnarray}
 At each time, $\xi_j(t)$ is a random number chosen with equal probability in the interval $(-\eta/2,\eta/2)$. Equivalently, Eq.~\eqref{eq2b} can be written as
\begin{eqnarray}
&& \mathbf{v}_j(t+1)=v_0  \mathcal{R}_\eta \Theta\!\left(\sum_{|\mathbf{x}_j-\mathbf{x}_i|<R_0}\mathbf{v}_j(t)-\beta\mathbf{x}_i(t)\right)\!, \label{eq2d}\\
&&\mathcal{R}_\eta= \begin{pmatrix} 
\cos\xi& -\sin\xi\\ 
\sin\xi&\cos\xi
\end{pmatrix}\!,\quad-\frac{\eta}{2}\leq\xi\leq\frac{\eta}{2}.\label{eq2e}
\end{eqnarray}
\end{subequations}
From now on, we redefine positions and velocities so that $v_0=1$ and $R_0/v_0$ replaces $R_0$. The results are Eqs.~\eqref{eq1} and \eqref{eq2} with $v_0=1$.
 
\section{Basic 2D exact solutions}\label{sec:3}
Now consider $d=2$ and $\eta=0$ (deterministic HCVM). Notice that the obtained 2D exact solutions are also exact solutions of the 3D deterministic HCVM provided all particles lie on the same plane. In the HCVM, arbitrarily many particles can occupy the same position and we can consider $G$ groupings with $N_j$ particles each, $\sum_{j=1}^G N_j=N$. The simplest case is $G=1$, all particles occupy the same site, move with the same velocities, and their positions and velocities coincide with those of the center of mass (CM). Then $\sum_{|\mathbf{x}_j-\mathbf{x}_i|<R_0}\mathbf{v}_j(t)=N\mathbf{W}(t)$, where $\mathbf{W}(t)$ is the CM velocity and $W(t)=|\mathbf{W}(t)|\in [0,1]$ is the {\em polarization} order parameter. 

\begin{figure}[ht]
\begin{center}
\includegraphics[clip,width=8cm]{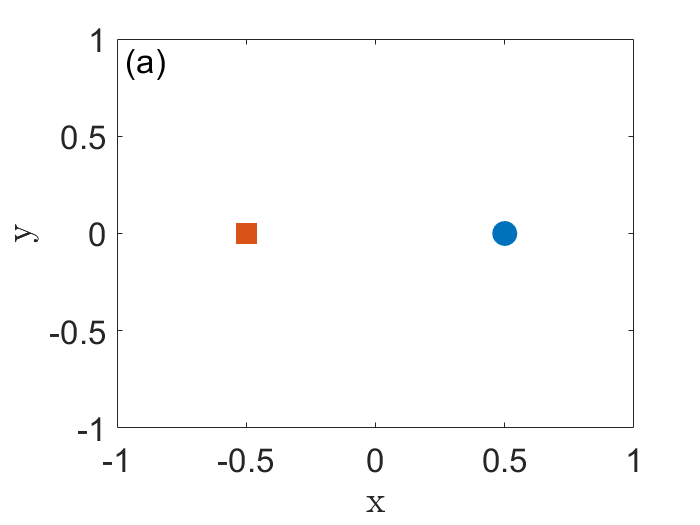}
\includegraphics[clip,width=8cm]{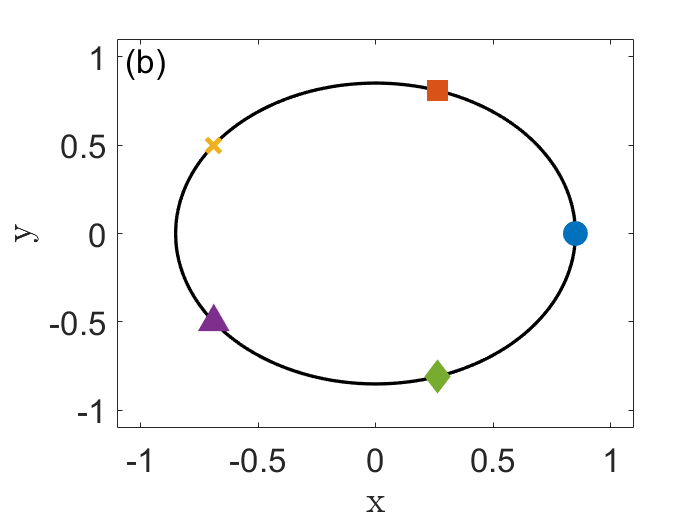}
\includegraphics[clip,width=8cm]{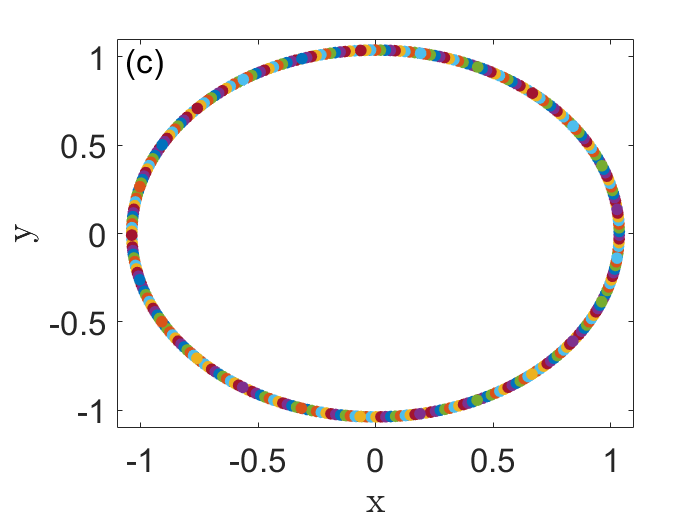}
\end{center}
\caption{(a) Positions of a period-2 exact solution for $\beta=4N$ and initial conditions $z(0)=\frac{1}{2}$, $v(0)=1$. (b) Same for a period-5 solution with $P=5$ in Eq.~\eqref{eq3c} and $z(0)=r_P$, $v(0)=e^{i\pi/P}$. (c) Same for $P=\sqrt{39}$. Colors distinguish positions at different times. \label{fig1}}
\end{figure}

\subsection{Invariant circles} We now use complex numbers instead of 2D vectors and consider that all particles occupy the same positions and move with the same velocities. We seek fully ordered solutions of Eqs.~\eqref{eq2} that are $P$-periodic for continuous time with positions proportional to $z^\frac{1}{P}$. After some algebra, we find 
\begin{subequations}\label{eq3}
\begin{eqnarray}
&&z(t\pm 1)=\frac{e^{\pm\frac{ i2\pi}{P}}e^{i\theta}}{2\sin\frac{\pi}{P}},\quad z(t)=\frac{e^{i\theta}}{2\sin\frac{\pi}{P}},\label{eq3a}\\ 
&&v(t+1)=ie^{i\theta}e^\frac{i\pi}{P},\quad v(t)=ie^{i\theta}e^{-\frac{i\pi}{P}},\label{eq3b}\\
&&\beta=\beta_N=4N\sin^2\frac{\pi}{P}, \quad r_P=\frac{1}{2\sin\frac{\pi}{P}}, \label{eq3c}
\end{eqnarray}
\end{subequations}
where $\theta$ is arbitrary. All $N$ particles are located at the same position at $t$, $x(t)=r_Pe^{i\theta}$, at a distance $r_P$ from the origin of coordinates, their speed and order parameter are 1, and the confinement parameter has the value of Eq.~\eqref{eq3c}. Then we have
\begin{eqnarray*}
z(t+1)-z(t)=e^{i\theta}\!\left[\frac{\cos\frac{2\pi}{P}-1+i\sin\frac{2\pi}{P}}{2\sin\frac{\pi}{P}}\right]\!=v(t+1),
\end{eqnarray*}
\begin{eqnarray*}
\sum_{|\mathbf{x}_j-\mathbf{x}_i|<R_0}v(t)-\beta z(t) = Ne^{i\theta}\!\left(ie^{-\frac{i\pi}{P}}-2\sin\frac{\pi}{P}\right)\\
=Ne^{i\theta}\!\left(i\cos\frac{\pi}{P}+\sin\frac{\pi}{P}-2\sin\frac{\pi}{P}\right)\!=Nie^{i(\theta+\frac{\pi}{P})} \\
\Longrightarrow\frac{\sum_{|\mathbf{x}_j-\mathbf{x}_i|<R_0}v(t)-\beta z(t)}{N}=v(t+1).
\end{eqnarray*}
Thus, Eqs.~\eqref{eq3} yield exact {\em invariant circles} of the positions in Eqs.~\eqref{eq2} for $\eta=0$ in 2D:
\begin{eqnarray}
z(k)=\frac{e^{i\theta}e^{\frac{i2k\pi}{P}}}{2\sin\frac{\pi}{P}},\,\, v(k)=ie^{i(\theta+\frac{(2k-1)\pi}{P})}\!,\,\, k\in\mathbb{Z}.    \label{eq4}
\end{eqnarray}
According to Eq.~\eqref{eq3c}, invariant circle solutions exist for $0<\beta\leq 4N$. These exact solutions are periodic if $P$ is a natural or a rational number and quasiperiodic if $P$ is irrational (both are $P$-periodic solutions in continuous time). After $P$ time steps, where $m$ is the integer closest to $P$, $x(t)$ is back at Eq.~\eqref{eq3a} with $\theta$ shifted by $\pm(P-m)$. For an irrational $P$, the resulting positions form a {\em quasiperiodic sequence} that will fill densely the invariant circle of radius $r_P$ given by Eq.~\eqref{eq3c}.  In fact, for any fixed value of the confinement parameter, there is an invariant circle solution of the HCVM given by Eq.~\eqref{eq4} with $P$ and radius:
\begin{equation}
P=\frac{\pi}{\arcsin\sqrt{\frac{\beta}{4N}}}, \quad r_P=\sqrt{\frac{N}{\beta}}. \label{eq5}
\end{equation}
The positions $z(t)$ change from one Riemann sheet of the function $r_P^{-1}z^\frac{1}{P}$ to the next one as time increases. Whenever $P=m$ with integer or rational $m$, the number of Riemann sheets is finite and the corresponding solution is a periodic one of the discrete time HCVM. Fig.~\ref{fig1} show the positions at different times for period-2, period-5 and quasiperiodic solutions of invariant circle type. We shall see later that the invariant circle solutions are stable for $\beta\leq 2N$ and for $\beta=4N$.

\begin{figure}[ht]
\begin{center}
\includegraphics[clip,width=8cm]{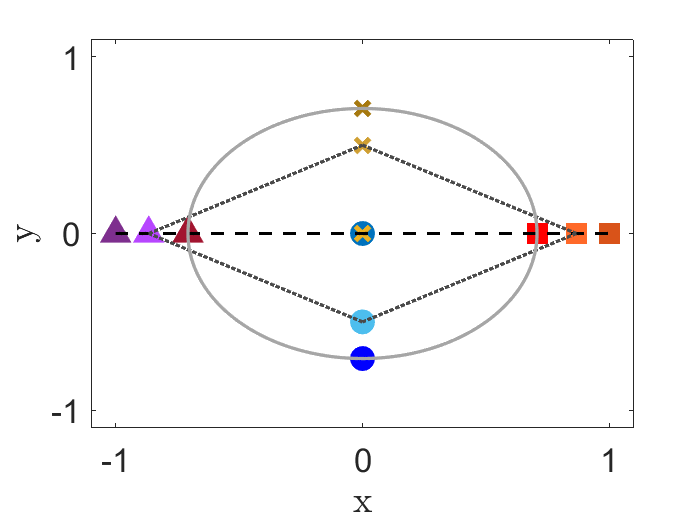}
\end{center}
\caption{Positions of a period-4 exact solutions for $\beta=4N$ (line with $z=0,\pm 1$, circle, cross, square and triangle), for $\beta=2N$ [rhombus with $z=\pm i/2,\pm\sqrt{3}/2$, and invariant circle Eq.~\eqref{eq4} with $P=4$]. Shapes and colors distinguish positions belonging to different solutions at different times. \label{fig2}}
\end{figure}

\subsection{Solutions on a line or rhombus}
The period 2 solutions of Fig.~\ref{fig1} lie on a straight line. We have found other fully ordered [$W(t)=1$] period 4 solutions that take positions on a straight line or a rhombus. For example, if $\beta=4N$ and initially we have $z(0)=0$, $v(0)=1$, we find the period-4 solution depicted in Fig.~\ref{fig2} with $z(1)=v(1)=1$, $z(2)=0$, $v(2)=-1$, $z(3)= v(3)=-1$. More generally, for $P=4$ and $\beta>N$, we have found solutions inside the unit circle with positions near $\pm e^{i\theta}$ and near 0,  and velocities  $\pm e^{i\theta}$:
\begin{subequations}\label{eq6}
\begin{eqnarray}
&&z(k)=e^{i\theta}\!\left(r+\sin\frac{k\pi}{2}\right)\!,\quad k\in\mathbb{Z},  \nonumber\\ %\label{eq6a}\\ 
&&\max\!\left\{-\frac{N}{\beta},\frac{N}{\beta}-1\right\}\!\leq r\leq\min\!\left\{\frac{N}{\beta},1-\frac{N}{\beta}\right\}\!,\nonumber\\
&&v(2k)=v(2k+1)=e^{i(\theta+k\pi)}. \label{eq6a}
\end{eqnarray}
Here $r$ and $\theta$ depend on the initial condition. We shall see later that these line solutions are stable if $\beta\geq 2N$. Similarly, for  $\beta>2N$, there exists the period-2 solution 
\begin{eqnarray}
&&z(k)=\!\left(r + \frac{1}{2}e^{ik\pi}\! \right)\!e^{i\theta},\quad \frac{N}{\beta}-\frac{1}{2}\leq r\leq\frac{1}{2}-\frac{N}{\beta},\nonumber\\
&&v(k)= e^{i(\theta+k\pi)}.\label{eq6b}
\end{eqnarray}\end{subequations}

There are other period-4 solutions that form a rhombus. One example is depicted in Fig.~\ref{fig2} for $\beta=2N$, with initial conditions $z(0)=-i/2$,  $v(0)=e^{-i\pi/6}$. Then $z(1)= \sqrt{3}/2$, $v(1)=e^{i\pi/6}$, $z(2)=i/2$, $v(2)=e^{i 5\pi/6}$, $z(3)=-\sqrt{3}/2$, $v(3)=e^{i 7\pi/6}$. The general form of the rhombus solution is
\begin{subequations}\label{eq7}
\begin{eqnarray}
&&z(2k)\!=\!e^{i\left[\theta+\left(k+\frac{1}{2}\right)\!\pi\right]}\!\sin\theta_0, \, z(2k\!+\!1)\!=\!e^{i (k \pi+\theta)}\!\cos\theta_0,\label{eq7a} \\
&&v(2k)\!=\!e^{i(k\pi+\theta+\theta_0)}, \, v(2k+1)=e^{i(k\pi+\theta-\theta_0)},\, k\in\mathbb{Z},\quad\label{eq7b}
\end{eqnarray}
\end{subequations}
which exists for $\beta=2N$. This solution interpolates between the period-4 solution Eq.~\eqref{eq6a} ($r=0$) found if $\sin(2\theta_0)=0$, and the period-4 circle solution Eq.~\eqref{eq4} ($P=4$) found if $|\sin\theta_0|=1/\sqrt{2}$. Fig.~\ref{fig2} depicts all these three exact solutions for $\theta=0$.

\section{Partially ordered and disordered solutions}\label{sec:4}
Eqs.~\eqref{eq6} are also exact deterministic periodic solutions if different particles have different values $r_j$ and $\theta_j$, $j=1,\ldots,N$, and $R_0$ is sufficiently large for all particles to form a single group. Partially ordered solutions of periods 2 and 4 have $|\sum_{j=1}^Ne^{i\theta_j}|/N=W<1$. Disordered solutions have $W=0$ and period 2. In fact, according to Eqs.~\eqref{eq2}, $v_j(t+1) = -1$ if $z_j(t)=1$. Next, $z_j(t+1)=0$ and therefore $v_j(t+2) = 1$ that implies $z_j(t+2)=1$, which is the initial situation. The same argument yields only two different positions and $P=2$ for any $z_j(t)=e^{i\theta}$ on the unit circle. 

As shown in \ref{ap:a}, for $\beta>NW$, we have found a partially ordered period-4 solution whose center of mass moves on a straight line with position and velocity:\begin{widetext}
\begin{subequations}\label{eq8}\begin{eqnarray}
X(k)= W e^{i\theta}\!\left(r+\sin\frac{k\pi}{2}\right)\!,\quad W(k)=We^{i\theta}\!\left(\sin\frac{k\pi}{2}+\cos\frac{k\pi}{2}\right)\!, \quad  k=0,1,\ldots, \label{eq8a}
\end{eqnarray}
respectively. The positions and velocities of the particles are
\begin{eqnarray}
&&z_j(k) = \frac{ \left(\frac{A_j}{2} + \sin\frac{k\pi}{2}\right)u_{1j} + i \left(B_j - 1 + e^{ik\pi}\right) \frac{u_{2j}}{2}}{B_ju_{1j} + iA_ju_{2j}} \frac{2NWe^{i\theta}}{\beta}, \nonumber\\
&& v_j(k) = \frac{ u_{1j} (\cos\frac{k\pi}{2} + \sin\frac{k\pi}{2}) + i u_{2j} e^{ik\pi}}{B_ju_{1j} + iA_ju_{2j}} \frac{2NWe^{i\theta}}{\beta}, \,\quad \label{eq8b}\\
&&u_{1j}=\pm\sqrt{\frac{\left(\frac{2NW}{\beta}\right)^2-A_j^2}{B_j^2-A_j^2}}, \quad u_{2j} = \pm \sqrt{\frac{B_j^2-\left(\frac{2NW}{\beta}\right)^2}{B_j^2-A_j^2}}, \label{eq8c}\\ 
&&\min\!\left\{-\frac{2NW}{\beta},B_j-2\right\}\!\leq A_j\leq\min\!\left\{\frac{2NW}{\beta},2-B_j\right\}\!, \, \frac{2NW}{\beta}\leq B_j\leq 2-A_j, \quad\nonumber
\end{eqnarray}
where the signs of $u_{2j}$ and $u_{1j}$, and the constants of $A_j$ and $B_j$ are chosen so as to satisfy the following constraints
\begin{eqnarray}
&&\sum_{j=1}^N\! u_{2j}e^{i\alpha_j}\!=0,\quad\frac{1}{N}\sum_{j=1}^N\! u_{1j}e^{i\alpha_j}\!=We^{i\theta}, \nonumber\\
&& \frac{1}{2N}\sum_{j=1}^N [A_j u_{1j} + i(B_j-1)u_{2j}] e^{i\alpha_j} \!= r We^{i\theta}, \nonumber\\
&& e^{i\alpha_j}\!=\frac{2NW e^{i\theta}/\beta}{B_ju_{1j} + iA_ju_{2j}}. \quad\label{eq8d}
\end{eqnarray}
\end{subequations}

\begin{figure}[ht]
\begin{center}
\includegraphics[clip,width=5.5cm]{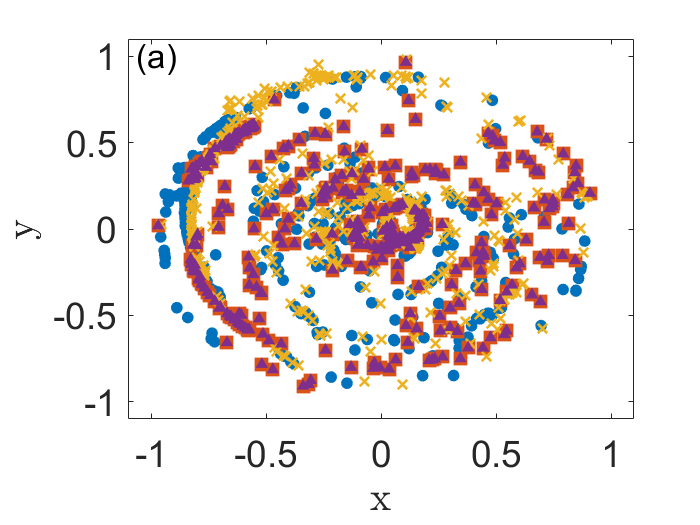}
\includegraphics[clip,width=5.5cm]{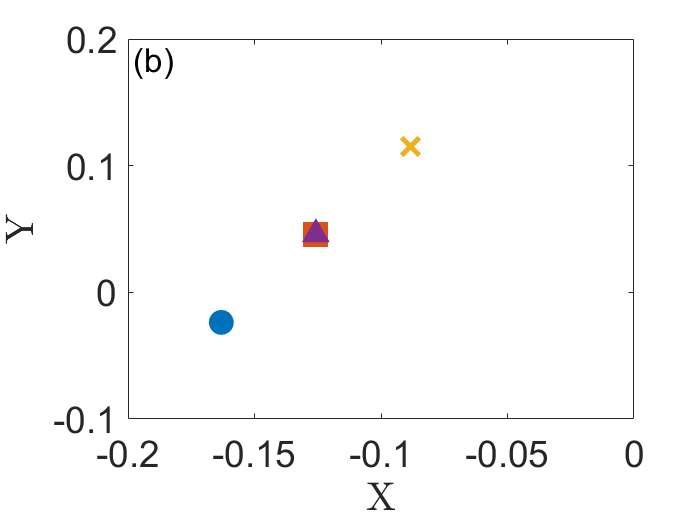}
\includegraphics[clip,width=5.5cm]{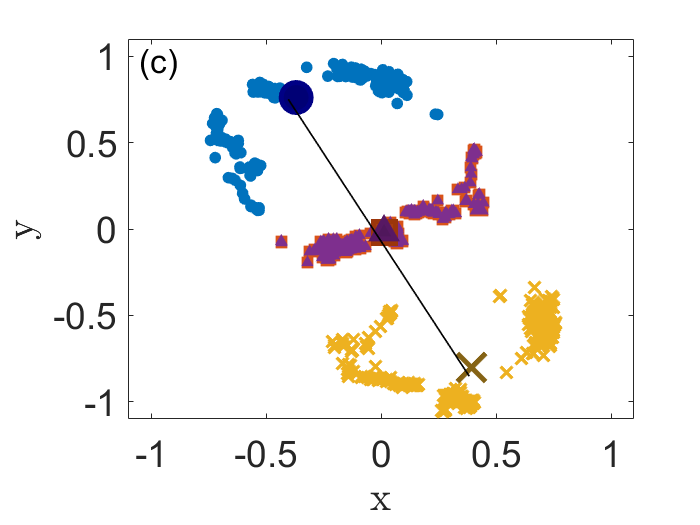}
\end{center}
\caption{Partially ordered period-4 solutions on a line. (a) Positions of the particles and (b) positions of the CM  for $R_0=2$. (c) Positions of the particles and positions of the CM on the depicted continuous line for $R_0=1$. Here $\beta=1500$ and $N=500$.  \label{fig3}}
\end{figure}

\begin{figure}[ht]
\begin{center}
\includegraphics[clip,width=5.5cm]{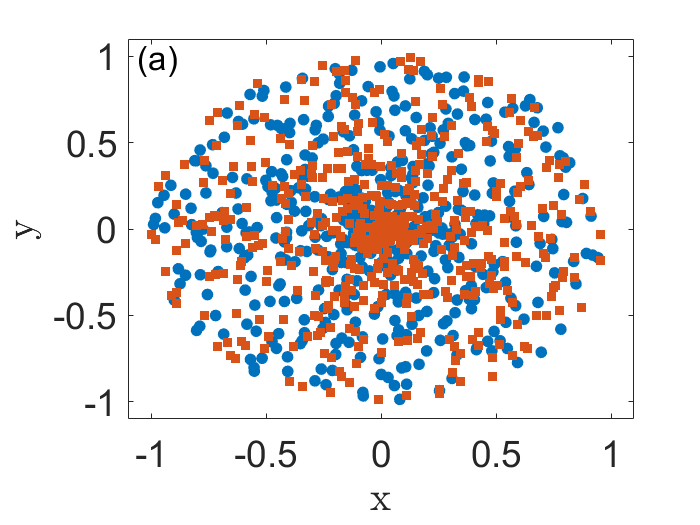}
\includegraphics[clip,width=5.5cm]{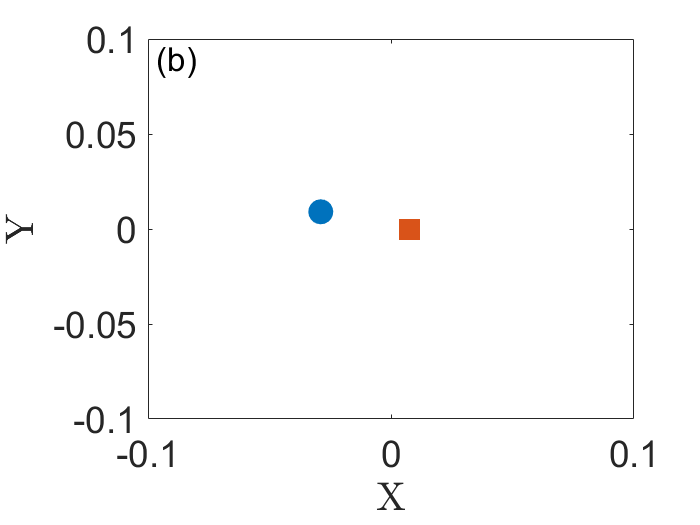}
\includegraphics[clip,width=5.5cm]{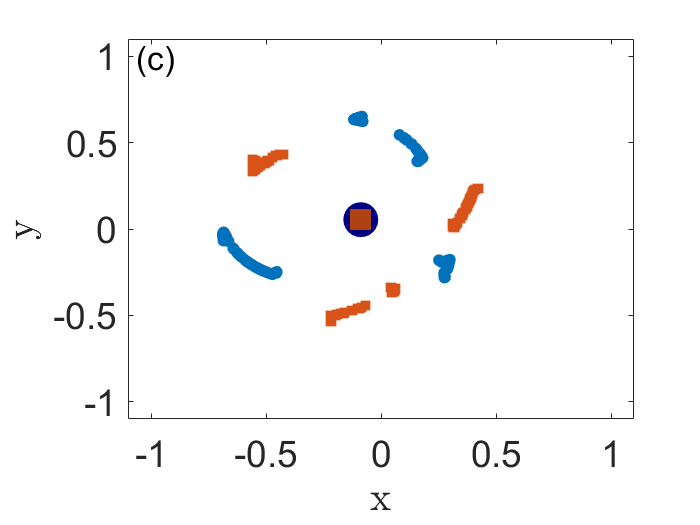}
\end{center}
\caption{Period-2 solutions for $N=500$. (a) Successive particles positions and (b) successive CM positions after $t=50000$ for a partially ordered exact solution corresponding to $\beta=10^9$, $R_0=2$ and random initial positions inside the unit circle with: $|\mathbf{x}_j(0)|<1$, $\mathbf{x}_j(0)=|\mathbf{x}_j(0)|\mathbf{v}_j(0)$, $\mathbf{v}_j(0)=(\cos\theta,\sin\theta)$, $-\pi<\theta<\pi$. (c) Same as (a) for the totally disordered solution ($W=0$) corresponding to $\beta=2000$, $R_0=1$.  The CM position is always at the origin. \label{fig4}}
\end{figure}
\end{widetext}
Fig.~\ref{fig3} depicts the 4-period exact solutions on a line given by Eqs.~\eqref{eq8}. The CM jumps between three values on a line. This is also the case if all particles have $u_{2j}=0$. However, if all particles have $u_{1j}=0$, then the CM has period 2 as in Figs.~\ref{fig4}(a) and \ref{fig4}(b).

If $u_{1j}=0$, then  $|A_{j}|=2NW/\beta$, $B_j-1=2r_j$, $|u_{2j}|=1$, $iu_{2j}e^{i\alpha_j}=e^{i\theta_j}$ and Eq.~\eqref{eq8b} becomes the partially ordered period-2 solution given by Eq.~\eqref{eq6b} with $r_j$ and $\theta_j$ instead of $r$ and $\theta$:
\begin{subequations}\label{eq9}
\begin{eqnarray}
z_j(k)=\!\left(r_j + \frac{1}{2}e^{ik\pi}\! \right)\!e^{i\theta_j},\quad v_j(k)= e^{i(\theta_j+k\pi)}.\label{eq9a}
\end{eqnarray}
 Then $0\leq W<1$; see Eq.~\eqref{eqa9b} in \ref{ap:a}. Fig.~\ref{fig4}(a) and \ref{fig4}(b) are the partially ordered period-2 solution whereas Fig.~\ref{fig4}(c) is the disordered period-2 solution. The restrictions Eq.~\eqref{eq8d} do not hold for period-2 solutions. 
 
 If $u_{2j}=0$, then Eqs.~\eqref{eq8} become the period-4 solution Eq.~\eqref{eq6a} with $r_j=A_j/2$ and $\theta_j$ instead of $\theta$:
 \begin{eqnarray}
z_j(k)\!=\!e^{i\theta_j}\!\!\left(\!r_j\!+\!\sin\!\frac{k\pi}{2}\!\right)\!\!,\, v_j(2k)\!=\!v_j(2k\!+\!1)\!=\!e^{i(\theta_j\!+\!k\pi)}\!. \quad\,   \label{eq9b}
\end{eqnarray}
\end{subequations}

\section{Several groups and orbit quantization}\label{sec:5}
The invariant circles given by Eqs.~\eqref{eq3} or \eqref{eq4} are not periodic for discrete time unless $P$ is a rational number $m+r/s$, where $m$, $r$ ($0\leq r<s$) and $s$ are integer numbers. When $r>0$, the period of the orbit is $ms+r$ (the period of the $s$th iterate of the map), whereas the period is $m$ when $P$ is an integer number $m$. Thus, for $P=m\in\mathbb{Z}$, the HCVM has quantized periodic orbits for confinement given by Eq.~\eqref{eq3c} which decreases as the period increases: $\beta$ is $4N$, $3N$, $2N$, $N$, for $P=2, 3, 4, 6$ and so on ($\beta=0$ for $P=1$). Can we lower the confinement parameter for specific values of $N$? 

\subsection{Partially polarized periodic orbits}
We can lower the confinement parameter by grouping particles in different sites given by Eq.~\eqref{eq4} provided the radius of influence $R_0$ is sufficiently small to encompass a single site. For example, if $N$ is even and $P=m$ is a natural number, we may split the swarm in two so that, at $t=0$, $N/2$ particles start at one of the $m$ roots of 1 and $N/2$ particles start at a different root with $\beta=\beta_N/2$, where $\beta_N$ is given by Eq.~\eqref{eq3c}. We obtain a $m$-periodic solution of Eqs.~\eqref{eq1} that has the same shape as the previous one. However, this solution will typically have a polarization smaller than 1. Clearly, we obtain  $m$-periodic solutions of Eqs.~\eqref{eq1} with decreasing $\beta$ up to $\frac{\beta_N}{m}$ provided $N$ is an integer multiple of $m$. For this last value of the confinement and provided all groupings have the same number of particles, the CM is always at the origin and all positions are equally occupied all the time, which implies the resulting solution to have a stationary profile with zero polarization.
\begin{figure}[ht]
\begin{center}
\includegraphics[clip,width=4cm]{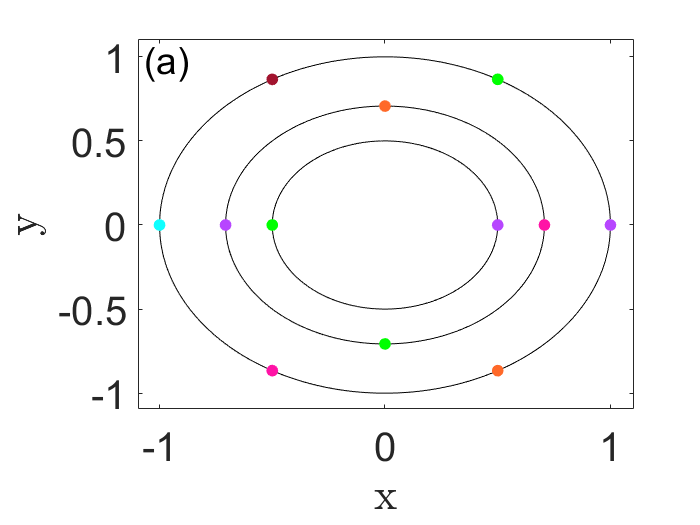}
\includegraphics[clip,width=4cm]{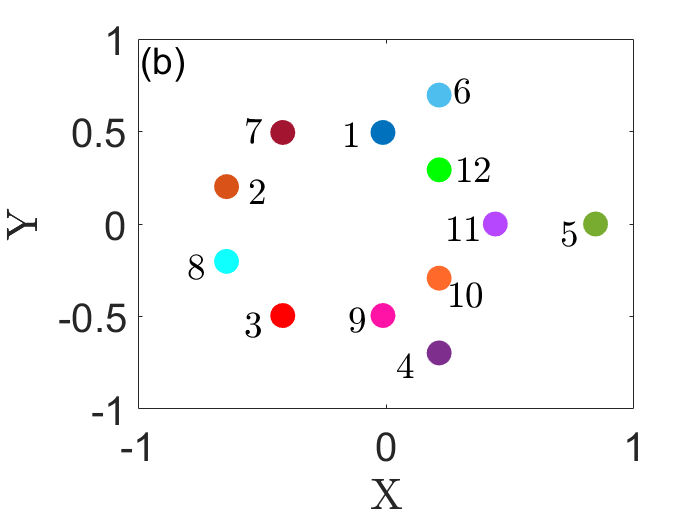}\\
\includegraphics[clip,width=4cm]{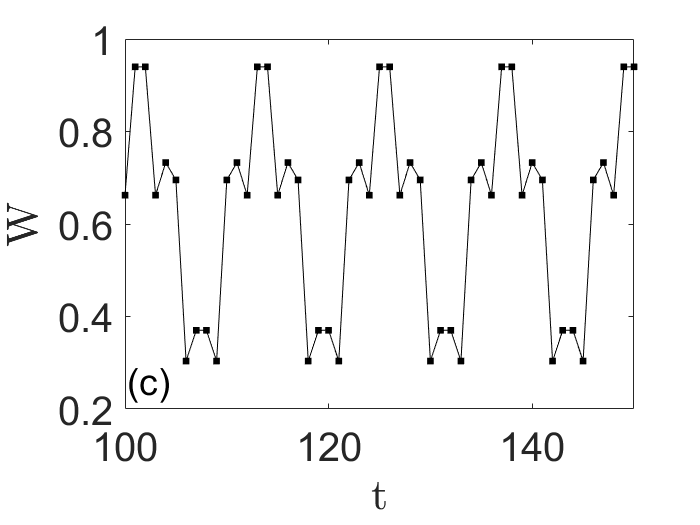}
\includegraphics[clip,width=4cm]{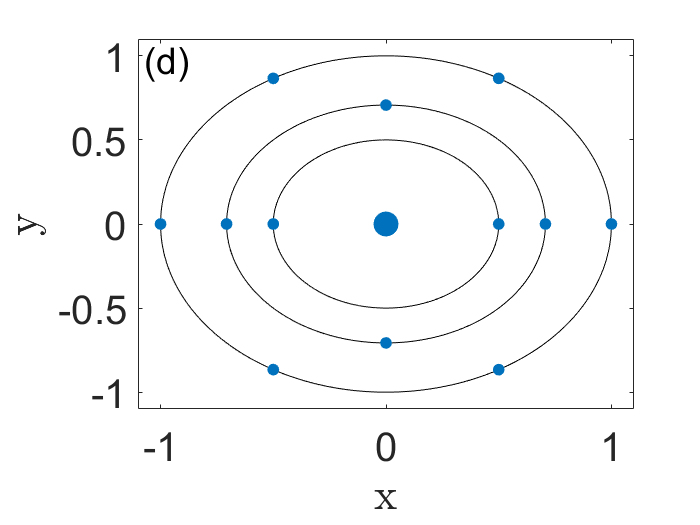}
\end{center}
\caption{Orbit quantization for a period-12 solution with $\eta=0$, $R_0=0.2$, $G=3$. Successive positions of (a) the particles and (b) the CM for $N=7$, $\beta=4$, $N_1=1$, $N_2=2$, $N_3=4$. The respective periods and radii of the orbits are $P_1=2$, $r_1=1/2$, $P_2=4$, $r_2=1/\sqrt{2}$, and $P_3=6$, $r_3=1$. Initial conditions for the groups: $z_j(0)=r_{j}$, $v_j(0)=ie^{-i\pi/P_j}$, with $j=1,2,3$. (c) Order parameter vs time for the same solution. (d) Stationary period-12 solution obtained by placing 1, 2 and 4 particles at each location of the inner, intermediate and outer orbits of Fig.~\ref{fig5}(a), respectively. The CM is at the origin, $W=0$, and $N=34$, $\beta=4$.  \label{fig5}}
\end{figure}

\begin{figure}[ht]
\begin{center}
\includegraphics[clip,width=4cm]{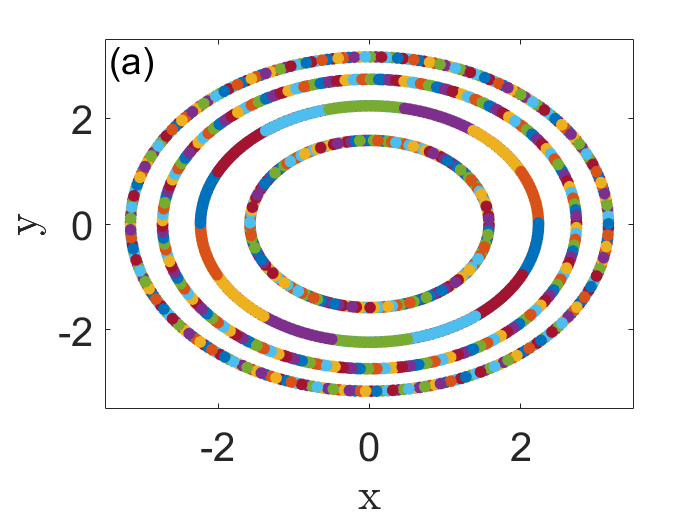}
\includegraphics[clip,width=4cm]{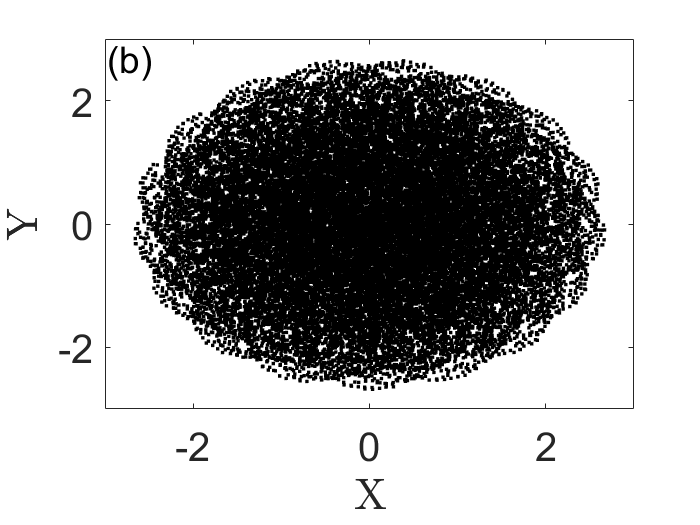}\\
\includegraphics[clip,width=4cm]{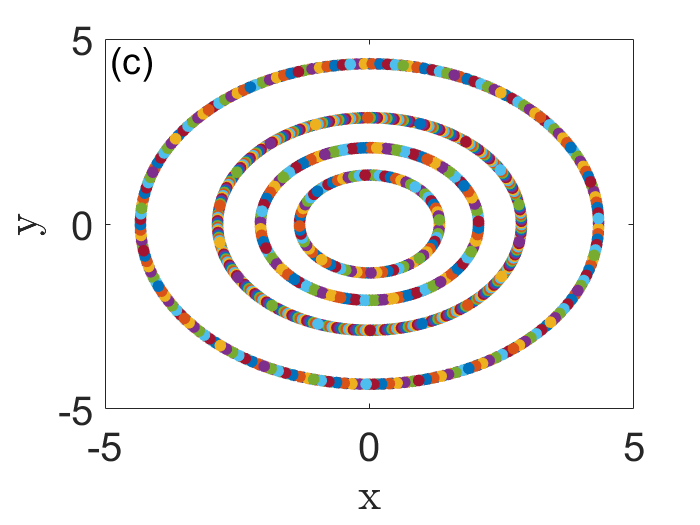}
\includegraphics[clip,width=4cm]{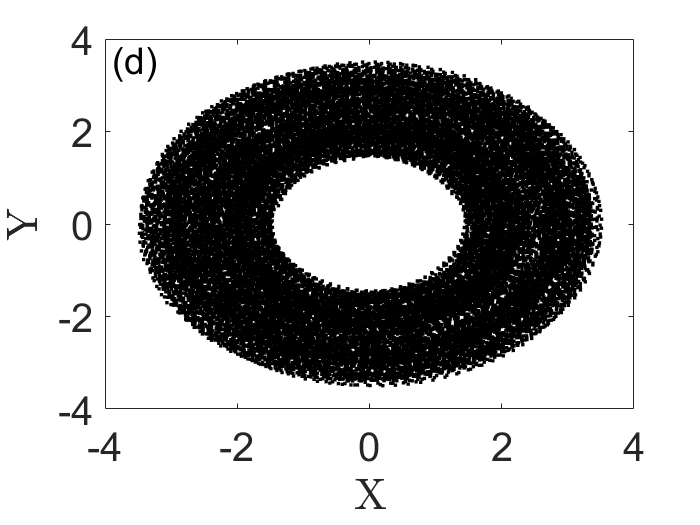}
\end{center}
\caption{(a) Concentric invariant circles and (b) the corresponding CM trajectory for $N_j=10j$, $j=1,2,3,4$, $N=100$. (c) and (d) are the same as (a) and (b) for  $N_1=7$, $N_2=17$, $N_3=33$, $N_4=75$ ($N=132$). Here $\beta=4$, $\eta=0$, $R_0=0.2$ and the common initial conditions for each group of particles are $z_j(0)=r_{Pj}$, $v_j(0)=ie^{-i\pi/P_j}$, $j=1,2,3,4$, where $P_j$ and $r_{Pj}$ are given by Eq.~\eqref{eq5}. The colors in (a) and (c) indicate different positions of each group of particles for different times. \label{fig6}}
\end{figure}

\subsection{Quantized periodic orbits}
Another possibility is to have periodic solutions forming concentric orbits with different radii depending on the value of $\beta$. For example, we can attain a period-12 solution with three groups ($G=3$) distributed in $N_1=\beta/4$ particles, period $P_1=2$, radius $r_1=1/2$; $N_2=\beta/2$, $P_2=4$, $r_4=1/\sqrt{2}$; $N_3=\beta$, $P_3=6$, $r_3=1$ if $\beta=4N/7$. The simplest case has $N=7$, $\beta=4$ (then $N_1=1$, $N_2=2$, $N_3=4$). The positions of the particles and of the CM are depicted in Figs.~\ref{fig5}(a) and \ref{fig5}(b), respectively, while Fig.~\ref{fig5}(c) shows the periodic evolution of the order parameter. By changing the number of particles in the groups, we can always have a stationary period-12 solution with the CM at the origin and $\beta/4$ particles occupy each of the 2 sites of the inner orbit, $\beta/2$ particles occupy each of the 4 sites of the intermediate orbit, and $\beta$ particles occupy each of the 6 sites of the outer orbit, as shown in Fig.~\ref{fig5}(d). It is clear that we can form more complicated high period solutions comprising concentric orbits by choosing appropriately $\beta$, groups, component periods and number of particles within each group. The resulting quantization of concentric orbits is reminiscent of Bohr's atomic model.

\subsection{Concentric quasiperiodic orbits}
What about concentric quasiperiodic orbits? To obtain four concentric orbits, we fix $\beta$ and the number of particles for each of the four groups. Then Eq.~\eqref{eq5} produces the (typically irrational) periods $P_j$ and radii $r_{Pj}$ of the four orbits, $j=1,2,3,4$. Fig.~\ref{fig6} shows two examples of concentric quasiperiodic orbits and the corresponding CM trajectories. Figs.~\ref{fig6}(a) and \ref{fig6}(b) correspond to $\beta=4$, $N_j=10j$, $j=1,2,3,4$, with $N=100$. Note that the CM seemingly fills a disk. For $N_1=7$, $N_2=17$, $N_3=33$ and $N_4= 75$ (so that $N=132$), we obtain Figs.~\ref{fig6}(c) and \ref{fig6}(d). The latter shows that the CM fills a corona region. It is clear that we can generalize these examples as desired.

\section{Stability}\label{sec:6}
Here we want to analyze the stability of the exact solutions we have found. We study their linear stability by means of the Floquet theorem proved here whereas  their nonlinear stability is explored in Section \ref{sec:7} by numerical simulations corresponding to different initial conditions. Given a solution of the 2D HCVM, the linearized equations about it are 
\begin{widetext}
\begin{subequations} \label{eq10}
\begin{eqnarray}
\delta\mathbf{\tilde{x}}_i(t+1)\!\!\!&=&\!\!\! \delta\mathbf{\tilde{x}}_i(t)+\delta\mathbf{\tilde{v}}_i(t+1),  \quad i=1,\ldots, N,  \label{eq10a}\\
\delta\mathbf{\tilde{v}}_i(t+1)\!\!\!&=&\! \!\!\mathcal{R}_\eta\!\left(\mathbb{I}_2-\frac{[\sum_{|\mathbf{x}_j-\mathbf{x}_i|<R_0}\mathbf{v}_j(t)-\beta\mathbf{x}_i(t)][\sum_{|\mathbf{x}_j-\mathbf{x}_i|<R_0}\mathbf{v}_j(t)-\beta\mathbf{x}_i(t)]^T}{|\sum_{|\mathbf{x}_j-\mathbf{x}_i|<R_0}\mathbf{v}_j(t)-\beta\mathbf{x}_i(t)|^2}\right)\!%\nonumber\\ &\cdot&\!\!
\cdot\!\frac{\sum_{|\mathbf{x}_j-\mathbf{x}_i|<R_0}\delta\mathbf{\tilde{v}}_j(t)-\beta\delta\mathbf{\tilde{x}}_i(t)}{|\sum_{|\mathbf{x}_j-\mathbf{x}_i|<R_0}\mathbf{v}_j(t)-\beta\mathbf{x}_i(t)|},\quad  \quad  \label{eq10b}
\end{eqnarray}\end{subequations}
where vectors are $2\times 1$ matrices and $A^T$ is the transpose of matrix $A$. 

\subsection{Periodic solutions and invariant circles}
Let us now consider the deterministic solutions Eqs.~\eqref{eq4} with particles distributed appropriately, either all in the same site as in Eq.~\eqref{eq3c} or in groups with a smaller value of $\beta$. Eqs.~\eqref{eq10} becomes
\begin{subequations}\label{eq11}
\begin{eqnarray}
&&\delta\mathbf{Y}_i(k+1)=\mathcal{A}(k)\delta\mathbf{Y}_i(k),\quad k=0,1,2,\ldots \label{eq11a}\\
&& \mathcal{A}(k)=\!\left(\begin{array}{cc} \mathbb{I}_2-\hat{\beta} A_k&A_k\\
-\hat{\beta} A_k&A_k\\ \end{array}\right)\!,\quad \delta\mathbf{Y}_i(t)=\!\left(\begin{array}{c} \delta\mathbf{\tilde{x}}_i(t)\\ \delta\mathbf{\tilde{v}}_i(t)\end{array}\right)\!,  \nonumber\\
&&
%\quad A_k=\!\left(\!\begin{array}{cc} \cos^2\varphi_k&\sin\varphi_k\cos\varphi_k\\ \sin\varphi_k\cos\varphi_k&\sin^2\varphi_k\\ \end{array}\!\right)\!, 
\quad\varphi_k=\theta+\frac{(2k-1)\pi}{P},\quad\psi_k=\theta+\frac{2k\pi}{P},\quad\hat{\beta}=\frac{\beta}{N}, \label{eq11b}\end{eqnarray}
where
\begin{eqnarray}
&&A_k=\frac{1}{S(\hat{\beta},P)}\!\left(\!\begin{array}{cc}
\cos^2\varphi_k-\!\left(1+\frac{\cos\psi_k\sin\varphi_k}{\sin\frac{\pi}{P}}\right)\!\hat{\beta}+\frac{\sin^2\psi_k}{4\sin^2\frac{\pi}{P}}\hat{\beta}^2&\sin\varphi_k\cos\varphi_k+\frac{\cos(\varphi_k+\psi_k)}{2\sin\frac{\pi}{P}}\hat{\beta}-\frac{\sin\psi_k\cos\psi_k}{4\sin^2\frac{\pi}{P}}\hat{\beta}^2\\
\sin\varphi_k\cos\varphi_k+\frac{\cos(\varphi_k+\psi_k)}{2\sin\frac{\pi}{P}}\hat{\beta}-\frac{\sin\psi_k\cos\psi_k}{4\sin^2\frac{\pi}{P}}\hat{\beta}^2& \sin^2\varphi_k+\!\left( \frac{\sin\psi_k\cos\varphi_k}{\sin\frac{\pi}{P}}-1\right)\!\hat{\beta}+\frac{\cos^2\psi_k}{4\sin^2\frac{\pi}{P}}\hat{\beta}^2\\
\end{array}\!\right)\!, \label{eq11c}\\
%&&A_k=\frac{1}{S(\hat{\beta},P)}\!\left(\!\begin{array}{c}
%\cos^2\varphi_k-\!\left(1+\frac{\cos\psi_k\sin\varphi_k}{\sin\frac{\pi}{P}}\right)\!\hat{\beta}+\frac{\sin^2\psi_k}{4\sin^2\frac{\pi}{P}}\hat{\beta}^2\\
%\sin\varphi_k\cos\varphi_k+\frac{\cos(\varphi_k+\psi_k)}{2\sin\frac{\pi}{P}}\hat{\beta}-\frac{\sin\psi_k\cos\psi_k}{4\sin^2\frac{\pi}{P}}\hat{\beta}^2 \end{array}\right. \nonumber\\
%&&\left.\quad\begin{array}{c} 
%\sin\varphi_k\cos\varphi_k+\frac{\cos(\varphi_k+\psi_k)}{2\sin\frac{\pi}{P}}\hat{\beta}-\frac{\sin\psi_k\cos\psi_k}{4\sin^2\frac{\pi}{P}}\hat{\beta}^2\\
%\sin^2\varphi_k+\!\left( \frac{\sin\psi_k\cos\varphi_k}{\sin\frac{\pi}{P}}-1\right)\!\hat{\beta}+\frac{\cos^2\psi_k}{4\sin^2\frac{\pi}{P}}\hat{\beta}^2\end{array}\!\right)\!, \label{eq11c}\\
&&S(\hat{\beta},P)=\left(1-\hat{\beta}+\frac{\hat{\beta}^2}{4\sin^2\left(\frac{\pi}{P}\right)}\right)^{\frac{3}{2}}. \label{eq11d}
\end{eqnarray}\end{subequations}\end{widetext}
For appropriate $\beta$ given by Eq.~\eqref{eq3c}, Eq.~\eqref{eq10} yields the simpler expression
\begin{subequations}\label{eq12}
\begin{eqnarray}
\delta\mathbf{\tilde{x}}_i(t\!+\!1)\!&=&\! \delta\mathbf{\tilde{x}}_i(t)+\delta\mathbf{\tilde{v}}_i(t+1),  \quad i=1,\ldots, N,  \label{eq12a}\\
\delta\mathbf{\tilde{v}}_i(t\!+\!1)\!&=&\!\! \left(\mathbb{I}_2\!-\!\mathbf{v}_i(t\!+\!1)\mathbf{v}_i(t\!+\!1)^T\right)\![\delta\mathbf{\tilde{v}}_i(t)\!-\!\beta\delta\mathbf{\tilde{x}}_i(t)] \quad\,\,\, \,\label{eq12b}\end{eqnarray}
Eq.~\eqref{eq11b} holds with  $\hat{\beta}=4\sin^2(\pi/P)$ and the matrix
\begin{eqnarray}
A_k=\!\left(\!\begin{array}{cc}\cos^2\varphi_{k+1}&\sin\varphi_{k+1}\cos\varphi_{k+1}\\
\sin\varphi_{k+1}\cos\varphi_{k+1}&\sin^2\varphi_{k+1}\\
\end{array}\!\right)\!. %=\frac{1}{2}\mathbb{I}+\frac{1}{2}\!\left(\!\begin{array}{cc}\cos(2\varphi_k)&\sin(2\varphi_k)\\\sin(2\varphi_k)&-\cos(2\varphi_k)\\
%\end{array}\!\right)\!,%label{eq13b}\\ &&\quad\varphi_k=\theta+\frac{(2k+1)\pi}{P}. 
 \label{eq12c}
\end{eqnarray}
\end{subequations}
When $P$ is a rational number, Eq.~\eqref{eq4} is a periodic solution. We shall consider the simpler case of positive integers $P>1$ that can be easily extended to rational $P$. The main stability result for a nonsingular matrix $\mathcal{A}(k)$ is the Floquet theorem, which can be derived as follows \cite{kel01}. Let $\Phi(t)$ be a fundamental nonsingular matrix solution of the linear Eqs.~\eqref{eq11a} for a generic particle $i$. Then $\Phi(k+P)$ is also a fundamental matrix and therefore,
\begin{subequations}\label{eq13}
\begin{eqnarray}
&&\Phi(k+P)= C\Phi(k),\quad C=\Phi(P)\Phi(0)^{-1}. \label{eq13a}\end{eqnarray}
The nonsingular matrix $C$ can be deduced from Eq.~\eqref{eq11a} with $k=P-1$ to be
\begin{eqnarray}
C=\mathcal{A}(P-1)\ldots \mathcal{A}(0),\label{eq13b}\end{eqnarray}
By inserting $\Phi(k)= B^kU(k)$ into Eq.~\eqref{eq13a}, we deduce the
\bigskip

\noindent
{\em  Floquet theorem.  The fundamental matrix $\Phi(k)$ can be written as the product of a nonsingular matrix $B^k$ times a periodic matrix of period $P$:}
\begin{eqnarray}
\Phi(k)= B^kU(k), \quad U(k+P)=U(k),\quad B^P=C,\quad\label{eq13c}
\end{eqnarray}\end{subequations}
{\em where $C$ is given by Eq.~\eqref{eq13a}.}\bigskip

\noindent Since $C$ is explicitly given by Eq.~\eqref{eq13b}, $B$ can be explicitly calculated from $B^P=C$ for integer or rational $P$. For a rational $P=m/n$, we redefine the period in Eqs.~\eqref{eq13} as the integer $m$.\bigskip

The trouble with using this result is that the matrices $\mathcal{A}(k)$ and $A_k$ are singular because they have zero determinants. The matrix $\mathcal{A}(k)$ has an eigenvalue 0 with eigenvector
\begin{eqnarray}
\mathbf{V}_0(k)=\! \left(\begin{array}{c}
0\\ 0\\ \sin\varphi_{k+1}\\-\cos\varphi_{k+1}\end{array}\right)\!.\label{eq14}
\end{eqnarray}
Eq.~\eqref{eq11a} implies that the fundamental matrix $\Phi(k)$ has zero determinant for any $k>0$ although we can form a nonsingular fundamental matrix $\Phi(0)$ by choosing the eigenvector $\mathbf{V}_0(0)$ as one of its columns. Then $\Phi(0)$ is invertible and we can define a matrix $C$ as in Eq.~\eqref{eq13a}, which will be singular and given by Eq.~\eqref{eq13b}. Since the solution of Eq.~\eqref{eq11a} with initial condition $\mathbf{V}_0(0)$ is the vector $(0,0,0,0)$ for all $k>0$,  the fundamental matrix $\Phi(k)$ has one column  with 0 in all its entries for $k>0$. The Floquet theorem \eqref{eq13c} could hold with a singular $B$. However, from Eq.~\eqref{eq13a} and $\Phi(k)=B^kU(k)$ we can only deduce $C[U(k+P)-U(k)]=0$ and the difference $U(k+P)-U(k)$ could be a rank 1 matrix having multiples of the eigenvector corresponding to the zero eigenvalue of $C$ in all its columns. One way out is requiring $U(k)$ to be $P$-periodic and orthogonal to the left eigenvector of $C$ for zero eigenvalue. We therefore have the following
\bigskip

\noindent
{\em  Floquet theorem. For a singular coefficient matrix $\mathcal{A}(k)$, the fundamental matrix $\Phi(k)$ can be written as the product of a singular matrix $B^k$ times a periodic matrix of period $P$ that is orthogonal to the left eigenvectors of the singular matrix $C$ of Eq.~\eqref{eq13a} corresponding to its zero eigenvalue:}
\begin{eqnarray}
\Phi(k)= B^kU(k), \quad U(k+P)=U(k),\quad B^P=C.\nonumber
\end{eqnarray}
\bigskip

\begin{figure}[ht]
\begin{center}
\includegraphics[clip,width=4cm]{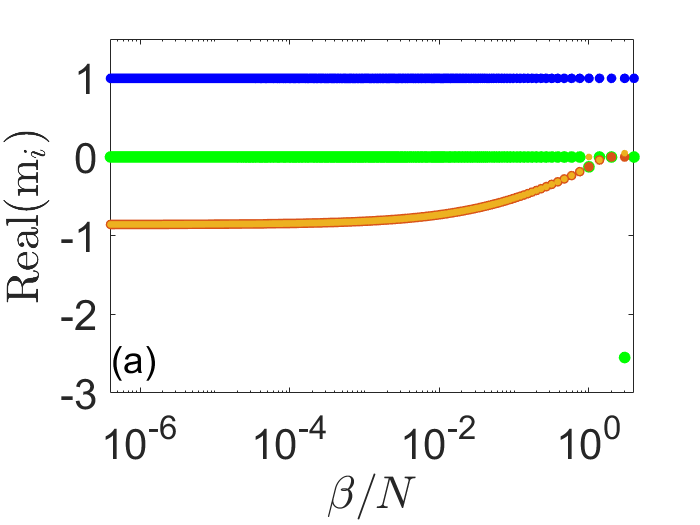}
\includegraphics[clip,width=4cm]{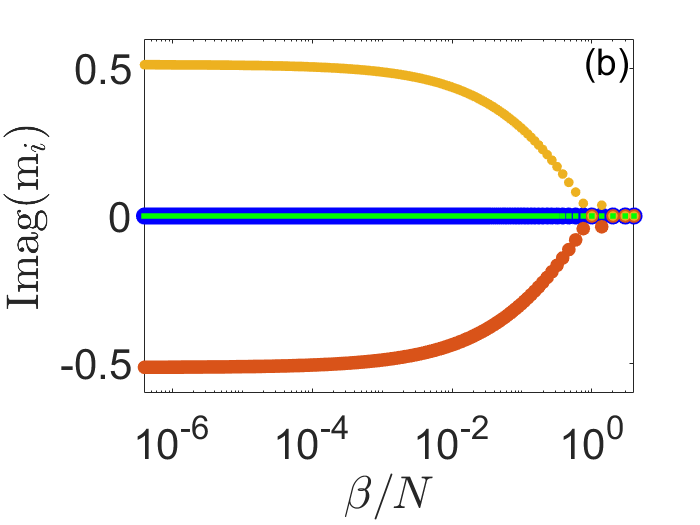}
\end{center}
\vspace{-5mm}
\caption{Real (a) and imaginary (b) parts of the Floquet multipliers $m_i$ for single orbit solutions with integer periods versus $\beta/N$. Panel (a) shows one Floquet multiplier with $|m_i|>1$ for $P=\frac{\beta}{N}=3$. \label{fig7}}
\end{figure}

The eigenvalues of the matrix $C$ are the Floquet multipliers and the solution \eqref{eq4} is unstable if at least one multiplier is outside the unit circle. One of the multipliers is always 0 because the matrix $\mathcal{A}(k)$ has one eigenvalue zero for any value of $P$. By performing an infinitesimal phase shift that changes $\theta$ to $\theta+\delta\theta$, $\delta\theta\ll 1$ in the invariant circle solution \eqref{eq4}, we observe that a vector function built from $z(k)$, $v(k)$ solves the linearized Eq.~\eqref{eq11a} or Eqs.~\eqref{eq12} with multiplier 1. Thus, the invariant circle solution \eqref{eq4} has two Floquet multipliers 0 and 1 and the other two Floquet multipliers decide whether the invariant circle is stable. Fig.~\ref{fig7} shows the real and imaginary parts of the Floquet multipliers of solutions with integer periods as a function of $\beta/N$. The multipliers different from 0 and 1 are complex conjugate, their moduli increase as $\beta/N$ decreases but they remain inside the unit circle. Thus, these periodic solutions are linearly stable.

For more complicated periodic solutions corresponding to concentric orbits as in Fig.~\ref{fig5}(a), the linearized coefficient matrix is larger: it is a $9\times 9$ matrix for that period 12 solution. Numerical calculation shows that the corresponding Floquet multipliers are 1 (triple) and 0 (sextuple).

Coming back to single orbit solutions, for a rational period $P=m/n$, the minimum period is $m$ with $C$ given by Eq.~\eqref{eq13b}. However, an integer multiple of $m$, $mp$, is also a period and the corresponding multipliers are the $p$th power of those of $C$. Multipliers outside the unit circle increase in absolute value whereas the absolute values of those inside the unit circle decrease. Thus, the linear stability of the solution with rational period is not affected. 

What about irrational periods $P$ corresponding to quasiperiodic invariant circles? Eq.~\eqref{eq13a} still holds but $C$ is no longer given by Eq.~\eqref{eq13b}. We may try to approximate $P$ by rational numbers $m/n$ with ever larger $m$ and plausibly {\em conjecture} that the stability of quasiperiodic invariant circle and periodic solutions with $P\approx m/n$ be the same. Sometimes $m$ is so large that the computation of the Floquet multipliers is not practical. The following construction seems to produce multipliers that preserve linear stability of the invariant circle although their values may not be quantitatively correct. Let $\hat{P}$ be the integer number closest to the irrational number $P$ and let $\mathcal{R}(P)$ be the rotation matrix of angle $\alpha_P=2\pi(P-\hat{P})/P$ needed to complete the period $P$ after $k=\hat{P}$ time steps. Then Eq.~\eqref{eq13b} has to be replaced by 
\begin{subequations}\label{eq15}
\begin{eqnarray}
&&C=\mathcal{A}(\hat{P}-1)\ldots \mathcal{A}(0)\mathcal{R}(P),  \label{eq15a}\\
&&\mathcal{R}(P)=\left(\!\begin{array}{cccc}
\cos\alpha_P&-\sin\alpha_P&0&0\\\sin\alpha_P&\cos\alpha_P&0&0\\0&0&\cos\alpha_P&-\sin\alpha_P\\0&0&\sin\alpha_P&\cos\alpha_P\\
\end{array}\!\right)\!.  \quad   \label{eq15b}
\end{eqnarray}
\end{subequations}
The rotated matrix $\mathcal{R}(P)C\mathcal{R}(P)^T$ has the same Floquet multipliers as $C$ and therefore $\mathcal{R}(P)$ can be written as the first factor before $\mathcal{A}(\hat{P}-1)$ instead of the last one in Eq.~\eqref{eq15a}.

\subsection{Periodic solutions on a line or rhombus}
The linearized equations about period-2 and period-4 exact solutions on a line or rhombus are Eqs.~\eqref{eq10}, which can be written in the form of Eqs.~\eqref{eq11a}-\eqref{eq11b} with a different expression for the matrix $A_k$. The Floquet theorem and Eqs.~\eqref{eq13a}-\eqref{eq13c} yield the Floquet multipliers for these solutions. Thus, calculating these multipliers we can discern whether these exact solutions are stable.

\section{Numerical results}\label{sec:7}
We have calculated the Floquet multipliers for the exact periodic and quasiperiodic solutions obtained in this work using the conjecture \eqref{eq15} for irrational periods. Except for the period-3 solution and invariant circles with $\beta>2N$, which are always unstable, all these solutions are linearly stable for appropriate values of $\beta$. In their range of stability, these solutions have a maximum Floquet multiplier of 1, the others are inside the unit circle. The unit Floquet multiplier appears because we can change arbitrarily the constant phase $\theta$ in our exact solutions.

\begin{figure}[ht]
\begin{center}
\includegraphics[clip,width=8cm]{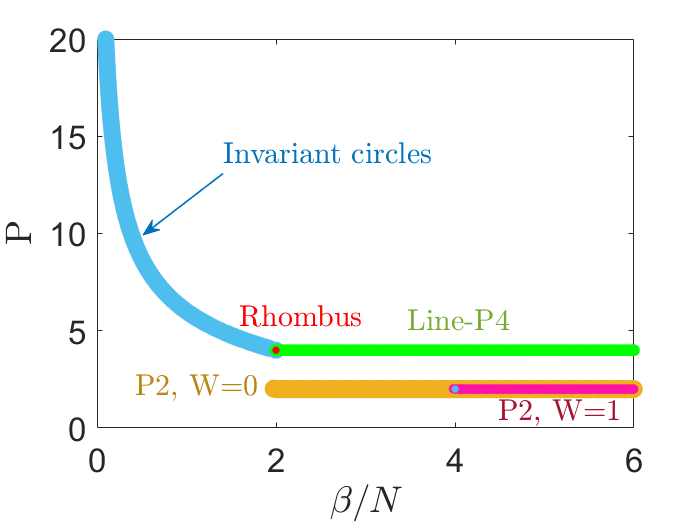}
\end{center}
\vskip-4mm
\caption{Stability for a single group of interacting particles. For $0<\beta\leq 2N$, the quasiperiodic (invariant circles) and periodic solutions of Eq.~\eqref{eq4} are stable (blue curve). For $\beta=4N$ the period-2 solution of this type is also stable. Period-4 solutions moving on a line, as in Eqs.~\eqref{eq6}, exist for $\beta>N$ and are stable for $\beta>2N$ and unstable for $N<\beta<2N$.  Period-2 solutions on a line exist for $\beta\geq 2N$ and are stable for $\beta\geq 4N$. The disordered period-2 solution is unstable for $0<\beta<2N$ and stable for $\beta>2N$. The rhombus solution exists and is stable for $\beta=2N$. \label{fig8}}
\end{figure}

\begin{figure}[ht]
\begin{center}
\includegraphics[clip,width=4cm]{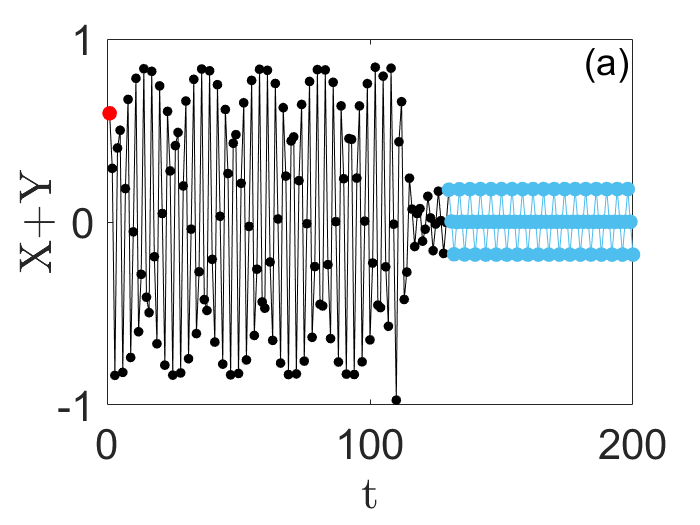}
\includegraphics[clip,width=4cm]{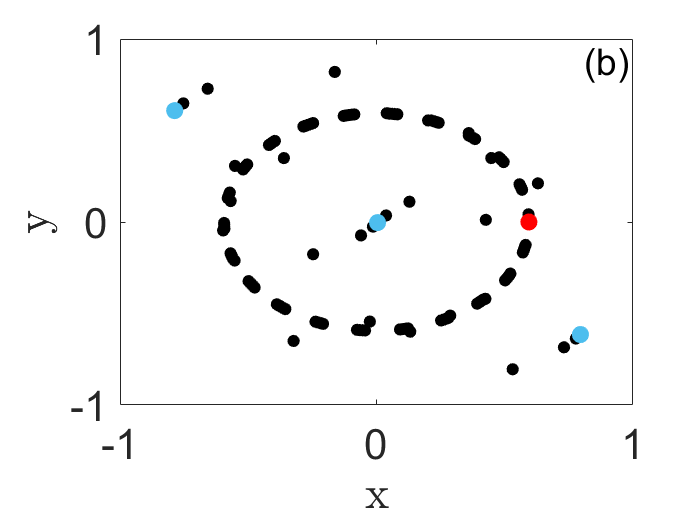}\\
\includegraphics[clip,width=4cm]{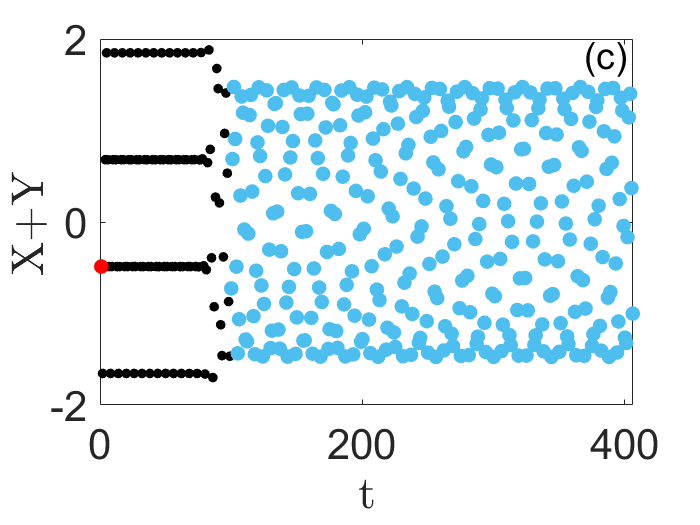}
\includegraphics[clip,width=4cm]{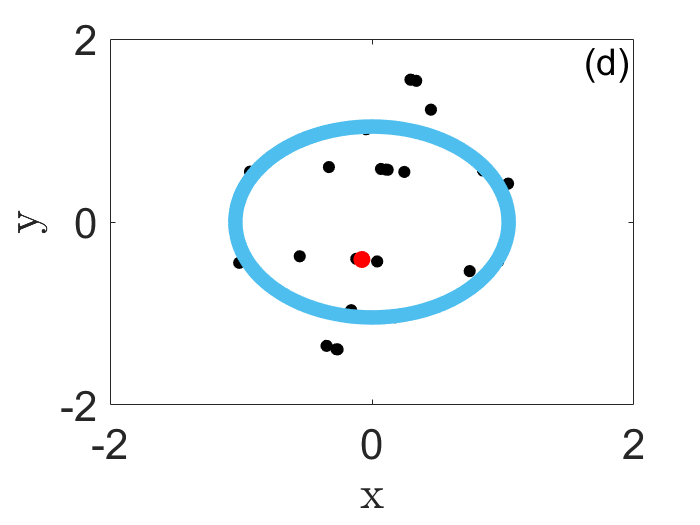}
\end{center}
\caption{Nonlinear stability of invariant circles with periods $P=\pi$, in panels (a) and (b), and $P=2 \pi$, in panels (c) and (d). Evolution of the center of mass coordinates $X(t)+Y(t)$ in panels (a) and (c) and $(X(t),Y(t))$ in panels (b) and (d) starting from the initial value marked by the red dot. The final values on the stable attractors are marked by blue dots and correspond to $\beta= 4N \sin^2 (\pi/P)$. \label{fig9}}
\end{figure}

\subsection{Linear stability}
Figure \ref{fig8} shows the linearly stable single-group solutions as a function of the parameter $\beta/N$. For $0<\beta\leq 2N$, the quasiperiodic (invariant circles) and periodic solutions of Eq.~\eqref{eq4} are linearly stable. For $2N\leq\beta<4N$, disordered period-2 solutions and ordered period-4 solutions moving on a line are linearly stable. For $\beta\geq 4N$, ordered and disordered period-2 solutions and ordered and partially ordered period-4 solutions are linearly stable. 

\subsection{Nonlinear stability}
The nonlinear stability of exact solutions depends on the confinement value and the initial condition. We distinguish three confinement intervals (i) $\beta>2N$, (ii) $\beta^*<\beta<2N$, (iii) $0<\beta<\beta^*$, where $\beta^*$ is the value below which chaotic solutions exist ($\beta^*\approx 0.2$ was numerically found in \cite{gon23mf}). 

For confinement interval (i), let us first consider the case of all particles occupying the same site initially. The final stable configuration is a period 2 or period 4 solution on the line, depending on the value of $r$ in Eqs.~\eqref{eq6}. If $|z(0)|<1$, $v(0)=z(0)/|z(0)|$, and $N/\beta < |z(0)| < 1 - N/\beta$, the final stable solution has period 2, whereas it has period 4 if $|z(0)| < N/\beta$ or $|z(0)| > 1- N/\beta$. If $|z(0)| > 1$, for example $n-1<|z(0)|<n$ where $n$ is an integer larger than 1, the final attractor has period 2 if $n-1+N/\beta < |z(0)| < n - N/\beta$ and it has period 4 if $|z(0)| < n-1+N/\beta$ or $|z(0)| > n - N/\beta$. When the particles occupy different sites randomly at $t=0$, the final solution is a partially ordered period 2 or period 4 solution under the conditions discussed in Section \ref{sec:4}.

For confinement interval (ii), any generic initial condition (randomly placed particles or all starting at a single site) evolves towards an invariant circle quasiperiodic solution with $\beta$ given by Eq.~\eqref{eq3c} or to a periodic solution for integer values of $P$ in Eq.~\eqref{eq3c}. Multiorbit solutions appear if the initial condition is close to the final linearly stable solution, as depicted in Figs.~\ref{fig5} or \ref{fig6} of Section \ref{sec:5}. 

The evolutions toward stable attractors from initial conditions close to invariant circles with irrational periods are shown in Fig.~\ref{fig9}. Fig.~\ref{fig9}(a)-(b) show how the unstable invariant circle with $P=\pi$ for $\beta= 4N\sin^21>2N$ in Eq.~\eqref{eq4}, region (i), evolves to the period-4 solution on a line. For $\beta= 4N\sin^2(1/2)\in(\beta^*,2N)$, region (ii), Fig.~\ref{fig9}(c)-(d) show the evolution to the stable invariant circle with $P=2\pi$ starting from a random initial position with $|z(0)|=r\leq 1$, and a random angle $\theta$ to be inserted into Eq.~\eqref{eq4} for $k=0$ and $r$ replacing $1/[2\sin(\pi/P)]$. These numerical solutions are fully consistent with the linear stability results depicted in Fig.~\ref{fig8}. We know that two of the Floquet multipliers are always 1 and 0. The other multipliers calculated using Eqs.~\eqref{eq15} are -2.59 and 0.028 for $P=\pi$ (therefore linearly unstable invariant circle) and $-0.14\pm 0.08 i$ for $P=2\pi$ (therefore linearly stable invariant circle). For $P=2\pi$, the rational approximations $P=6.3, 6.28, 6.283, 6.2832, 6.28319$ produce Floquet multipliers 1 and 0 (triple) whereas the integer approximation $P=6$ yields -0.125 (double), 1, 0. For the unstable solution with period $P=\pi$, the rational approximation $P=3.1$ produce Floquet multipliers -18600 and 0, whereas better rational approximations $P=3.14, 3.142, 3.1416, 3.14159$ yield much larger multipliers outside the unit circle. Of course, Eqs.~\eqref{eq15} yield multipliers -2.55, 0.05 for all approximations, thereby indicating instability of the quasiperiodic solution. Similar discrepancies appear when calculating successive approximations to unstable solutions with rational periods, for example $P=8/3$. The exact multipliers are 6.6 and 0.1 in addition to 0 and 1. The approximations $P=3$ and $P=2.6=13/5$, have different multipliers $-2.55; 0.05$ and $-15.44; 0.0015$, but both indicate that the corresponding exact periodic solution is unstable. 

For multiorbit solutions with irrational periods, the coefficient matrices of the linearized equations are larger but we can use Eq.~\eqref{eq15a} with integer period closest to the sum of periods of the orbits and a larger rotation matrix with blocks given by Eq.~\eqref{eq15b}. For the case of Fig.~\ref{fig6}(a), we obtain the 16 multipliers $0.0197\pm0.0208 i$, 0.0012, and 0 (13-ple), which suggests this solution to be stable. For Fig.~\ref{fig6}(b), the multipliers are $0.9416\pm 1.1496 i$ and 0 (14-ple), suggesting instability. 

\begin{figure}[ht]
\begin{center}
\includegraphics[clip,width=4cm]{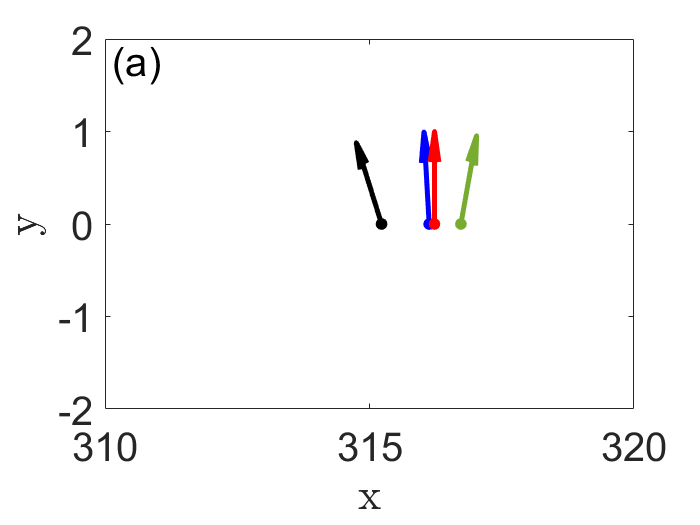}
\includegraphics[clip,width=4cm]{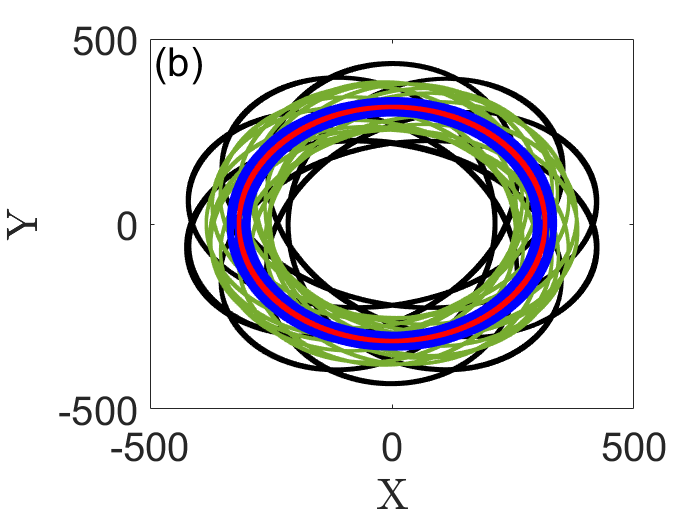}\\
\includegraphics[clip,width=4cm]{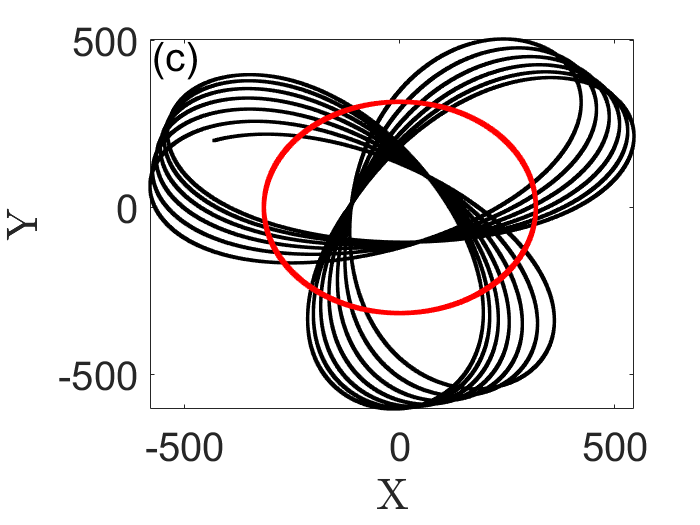}
\includegraphics[clip,width=4cm]{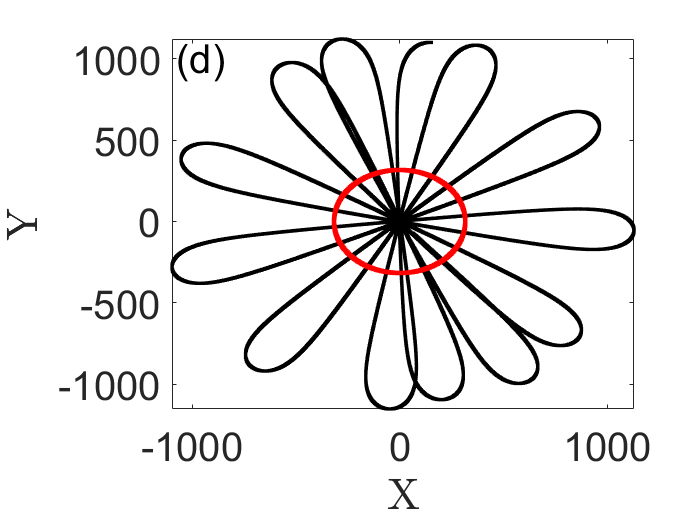}
\end{center}
\caption{Trajectories from initial conditions that depart from perturbed invariant circles for $\beta=10^{-5}<\beta^*$. All particles are occupy the same site, which is equivalent to having $N=1$. (a) Initial velocities (the red arrow marks initial conditions  $z(0)=r_P$, $v(0)=i e^{-i\pi/P}$ on the invariant circle, the other arrows have $ z(0) = r_P + \epsilon$, $v(0) = e^{i(\delta-\pi/P)}$ with blue: $\epsilon = - 0.1$, $\delta = 0.1$, green: $ \epsilon = 0.5$, $\delta = - 0.3$, black: $\epsilon = - 1$, $\delta = 0.5$), and (b) the corresponding trajectories moving on an annulus about the invariant circle. (c), (d) The initial conditions are (c) $z(0) = r e^{ ia}$, $v(0) = e^{ib}$ with $r$ randomly chosen in $(0,2)$ ($r=0.8340$, $a=4.5259$, $b=0.0007$), and (d) $z(0) = r e^{i a}$, $v(0) = e^{ia}$, $r$ randomly chosen in $(0,1)$ ($r=0.4170$, $a=4.5259$). $a$, $b$ are randomly chosen in $(0,2\pi)$. The invariant circle is depicted in red color. In all cases, 30000 iterations have been depicted after a transient period of de $5\times 10^5$ iterations. \label{fig10}}
\end{figure}

For confinement interval (iii), $\beta<\beta^*$, let us first consider the case in which all the particles occupy the same site. Then the HCVM is the deterministic mean field model because the 3D trajectories always move on a plane and are the same as found in the 2D HCVM; see Figure 3 of \cite{gon23mf} and compare it to Fig.~\ref{fig10} below. If the initial condition is compatible with Eqs.~\eqref{eq3}, the numerical solution produces the invariant circle. If the initial condition is a small disturbance of Eqs.~\eqref{eq3}, the particles trajectory fills an annulus whose width depends on $(\beta^*-\beta)$ and on the disturbance size as shown in Fig.~\ref{fig10}(b). For initial conditions further away from the invariant circle, the trajectory presents several loops as depicted in Figs.~\ref{fig10}(c) and \ref{fig10}(d). All these attractors are chaotic with a largest Lyapunov exponent about $4\times 10^{-5}$. These patterns are the same as depicted in Figure 3(a) of \cite{gon23mf}; see Fig.~3(b) of \cite{gon23mf} for the multifractal singularity spectrum for the transition between quasiperiodicity and chaos. We have observed excitability in Region (iii). For $\beta$ close to $\beta^*$ in region (iii), patterns as in Figs.~\ref{fig10} return to the invariant circle solution after some time. This time becomes longer and longer the smaller $\beta$ becomes until the are seemingly permanent for  very small $\beta$. Thus,  the HCVM behaves as an excitable system. Small departures from the invariant circle return to it rapidly, larger departures perform a long excursion through the patterns of Fig.~\ref{fig10}. For the case of random initial conditions for $N$ particles, patterns like those in Fig.~\ref{fig10} appear and become permanent for fixed $\beta$ and sufficiently large $N$. More complex chaotic attractors may appear that involve several groups of particles moving on irregular non circular orbits. These phenomena will be explored in future work.

\section{Conclusions}\label{sec:8}
We have found different exact periodic and quasiperiodic solutions of the 2D or 3D harmonically confined Vicsek model without alignment noise, analyzed their linear stability for the 2D case and compared the results to direct numerical simulations. For specific quantized values of the confinement strength, there are periodic solutions with integer or rational periods and quasiperiodic solutions (invariant circles) that have irrational periods. Many or all particles may occupy the same site at each time, have up to unit polarization, and move on a single or on concentric circular orbits. There are also period 2 and period 4 solutions on a line for a range of confinement strengths and period 4 solutions on a rhombus. These solutions may be fully ordered (unit polarization), but there are partially ordered period 4 solutions and totally disordered (zero polarization) period 2 solutions. All polarized solutions of the HCVM move within a bounded region, in contrast to the migratory flocking solutions of the VM  with periodic boundary conditions \cite{vic95,cha20} and of other models \cite{bir07,cuc07,has13,erb16}. We have explored the linear stability of the exact periodic solutions in two dimensions using the Floquet theorem (extended to the case of singular coefficient matrices) and verified the stability assignments by direct numerical simulations. For invariant circles corresponding to irrational periods, we have used the conjectured Eqs.~\eqref{eq15} and found that their linear stability is the same as that of the periodic orbit of integer period closest to the irrational number.

Future works may explore excitability phenomena in the 2D HCVM, the stability of the exact solutions in 3D and transitions to chaotic attractors. Ascertaining the effect of alignment noise on these exact solutions is an open problem.
\bigskip

{\em Acknowledgments.} We thank Bj\"orn Birnir for suggesting that we search for exact solutions of the HCVM. This work has been supported by the FEDER/Ministerio de Ciencia, Innovaci\'on y Universidades -- Agencia Estatal de Investigaci\'on (MCIN/ AEI/10.13039/501100011033) grant  PID2020-112796RB-C22, by the Madrid Government (Comunidad de Madrid-Spain) under the Multiannual Agreement with UC3M in the line of Excellence of University Professors (EPUC3M23), and in the context of the V PRICIT (Regional Programme of Research and Technological Innovation). 

\appendix
\section{Partially ordered solution} \label{ap:a}
We try the exact solution:
\begin{subequations}\label{eqa1}
\begin{eqnarray}
&&z_j(k)=r_je^{i\theta_j} + \!\left(u_{1j}\sin\frac{k\pi}{2} + i \frac{u_{2j}}{2}e^{ik\pi}\right)\! e^{i\alpha_j},\label{eqa1a}\\
&&v_j(k)=\!\left[u_{1j}\!\left(\cos\frac{k\pi}{2}+\sin\frac{k\pi}{2}\right)\! +i u_{2j}e^{i k\pi}\right]\! e^{i\alpha_j},\quad\label{eqa1b}\\
&&u_{2j}^2+u_{2j}^2=1, \label{eqa1c}
\end{eqnarray}
where $u_{1j}$ and $u_{2j}$ are real, $|v_j(k)|=1$,
\begin{eqnarray}
&&\frac{1}{N}\sum_{j=1}^Nu_{1j}e^{i\alpha_j}=\frac{1}{N}\sum_{j=1}^Nv_j(k)=W\,e^{i\theta}, \label{eqa1d} \\ 
&& \sum_{j=1}^N u_{2j}e^{i\alpha_j}=0,\label{eqa1e}\\
&&\frac{1}{N}\sum_{j=1}^Nr_j e^{i\theta_j}= rW e^{i\theta},  \label{eqa1f}
\end{eqnarray}
and $W$ is the order parameter. These equations modify the period-4 solution Eq.~\eqref{eq6a}. Then the positions and velocities of the center of mass are:
\begin{eqnarray}
&&X(k)= W e^{i\theta}\!\left(r+\sin\frac{k\pi}{2}\right)\!,\nonumber\\
&& W(k)=We^{i\theta}\!\left(\sin\frac{k\pi}{2}+\cos\frac{k\pi}{2}\right)\!, \label{eqa1g}
\end{eqnarray}\end{subequations}
respectively. Note that $W(0)=W(1)=W e^{i\theta}=-W(2)=-W(3)$. Then Eq.~\eqref{eq2d} with $\eta=0$ produces for $k=1,3$:
\begin{subequations}\label{eqa2}
\begin{eqnarray}
&&u_{1j}-iu_{2j}=\frac{W_0 e^{i(\theta-\alpha_j)}-r_j e^{i(\theta_j-\alpha_j)}-\frac{i}{2}u_{2j} }{a_j},\quad\label{eqa2a}\\
&&a_j=\!\left|W_0 e^{i(\theta-\alpha_j)}-r_j e^{i(\theta_j-\alpha_j)}-\frac{i}{2}u_{2j}\right|\!,\label{eqa2b}\\
&&u_{1j}+iu_{2j}=\frac{W_0 e^{i(\theta-\alpha_j)}+r_j e^{i(\theta_j-\alpha_j)}+\frac{i}{2}u_{2j} }{c_j},\label{eqa2c}\\
&&c_j=\left| W_0 e^{i(\theta-\alpha_j)}+r_j e^{i(\theta_j-\alpha_j)}+\frac{i}{2}u_{2j}\right|\!,\label{eqa2d}\\
&& W_0=\frac{NW}{\beta}, \label{eqa2e}
\end{eqnarray}\end{subequations}
whereas for $k=2,4$, we get
\begin{subequations}\label{eqa3}
\begin{eqnarray}
&&u_{1j}\!-\!iu_{2j}\!=\!\frac{-W_0 e^{i(\theta-\alpha_j)}\!\!+\!r_j e^{i(\theta_j-\alpha_j)}\!\!+\!u_{1j}\!-\!\frac{i}{2}u_{2j} }{b_j},\quad\quad\label{eqa3a}\\
&&b_j=\!\left|W_0 e^{i(\theta-\alpha_j)}-r_j e^{i(\theta_j-\alpha_j)}-u_{1j}+\frac{i}{2}u_{2j}\right|\!,\label{eqa3b}\\
&&u_{1j}\!+\!iu_{2j}\!=\!\frac{-W_0 e^{i(\theta-\alpha_j)}\!\!-\!r_j e^{i(\theta_j-\alpha_j)}\!\!+\!u_{1j}\!+\!\frac{i}{2}u_{2j} }{d_j},\label{eqa3c}\\
&&d_j=\left| W_0 e^{i(\theta-\alpha_j)}+r_j e^{i(\theta_j-\alpha_j)}-u_{1j}-\frac{i}{2}u_{2j}\right|\!.\label{eqa3d}
\end{eqnarray}\end{subequations}
Adding the numerators of Eqs.~\eqref{eqa3a} and \eqref{eqa2a}, we obtain $(a_j+b_j)(u_{1j}-iu_{2j})= u_{1j}-iu_{2j}$, from which $a_j+b_j=1$.  A similar relation follows from Eqs.~\eqref{eqa2c} and \eqref{eqa3c}. Thus. we have
\begin{eqnarray}
a_j+b_j=1,\quad c_j+d_j=1,\quad a_j, b_j, c_j, d_j\in(0,1).  \quad     \label{eqa4}
\end{eqnarray}

Adding Eqs.~\eqref{eqa2a} and \eqref{eqa2c}, 
\begin{eqnarray*}
2W_0e^{i(\theta-\alpha_j)}=(a_j+c_j)u_{1j}+i(c_j-a_j)u_{2j},
\end{eqnarray*}
which yields
\begin{eqnarray}
e^{i\alpha_j} = \frac{2 W_0 e^{i\theta}}{(a_j + c_j) u_{1j} + i \left(c_j - a_j\right) u_{2j}}. \label{eqa5}
\end{eqnarray}
Using now Eq.~\eqref{eqa1c} and $|e^{i\alpha_j}|=1$, we get
\begin{subequations}\label{eqa6}
\begin{eqnarray}
u_{1j}=\pm\sqrt{\frac{4W_0^2-(c_j-a_j)^2}{4a_jc_j}}, \label{eqa6a}\\
u_{2j}=\pm\sqrt{\frac{(a_j+c_j)^2-4W_0^2}{4a_jc_j}}, \label{eqa6b}\\
|c_j-a_j|\leq 2W_0\leq c_j+a_j. \label{eqa6c}
\end{eqnarray}\end{subequations}
Subtracting Eq.~\eqref{eqa2a} from \eqref{eqa2c} and using Eq.~\eqref{eqa1c}, we obtain
\begin{subequations}\label{eqa7}
\begin{eqnarray}
&&r_je^{i\theta_j}=\frac{1}{2}e^{i\alpha_j}[(c_j-a_j)u_{1j}+i(c_j+a_j-1)u_{2j}],\quad \label{eqa7a}\\
&&r_je^{i\theta_j}=\frac{e^{i\theta}}{4W_0}\!\left[\frac{W_0^2(c_j-a_j)}{a_jc_j}+(4a_jc_j-a_j-c_j)\right.\nonumber\\
&&\quad\quad\,\,\left. \times\left(\frac{c_j^2-a_j^2}{4a_jc_j}+iu_{1j}u_{2j}\right)\right]\!. \label{eqa7b}
\end{eqnarray}
\end{subequations}
We now define
\begin{subequations}\label{eqa8}
\begin{eqnarray}
&&A_j=c_j-a_j,\, B_j=c_j+a_j,\, |A_j|\leq\frac{2NW}{\beta}\leq B_j, \quad\label{eqa8a}\\
&&0<A_j+B_j, \quad0<B_j-A_j.    \label{eqa8b}
\end{eqnarray}\end{subequations}
Then Eqs.~\eqref{eqa6} become Eq.~\eqref{eq8c}, Eqs.~\eqref{eqa1a}-\eqref{eqa1c} and \eqref{eqa7a} yield Eq.~\eqref{eq8b}, and Eqs.~\eqref{eqa1d}-\eqref{eqa1f} with Eq.~\eqref{eqa7a} are the constraints Eq.~\eqref{eq8d}. Eq.~\eqref{eqa4} implies that $0<B_j<2$.

If $u_{1j}=0$, $|A_j|=2W_0$, and $u_{2j}=\pm 1$. Then Eq.~\eqref{eqa7a} implies 
\begin{subequations}\label{eqa9}
\begin{eqnarray}
r_j=\frac{B_j-1}{2},\quad\alpha_j=\theta_j-\frac{\pi}{2}+\pi(1+u_{2j}).\label{eqa9a}
\end{eqnarray}
and Eqs.~\eqref{eqa1a} and \eqref{eqa1b} become
\begin{eqnarray}
z_j(k)=\!\left(r_j+\frac{1}{2} e^{ik\pi}\right) e^{i\theta_j},\quad v_j(k)=e^{i(k\pi+\theta_j)}, \quad\label{eqa9b}
\end{eqnarray}
\end{subequations}
which generalize Eq.~\eqref{eq6a} to a partially ordered period-2 solution with $0\leq W<1$.

\end{document}